\documentclass[fleqn,usenatbib]{mnras}

\usepackage{newtxtext,newtxmath}

\usepackage[T1]{fontenc}
\usepackage{ae,aecompl}

\usepackage{graphicx}	
\usepackage{amsmath}	
\usepackage{amssymb}	
\usepackage{romannum}

\title[\textsc{Neural network based reionization constraints}]{Reionization history constraints from neural network based predictions of high-redshift quasar continua}

\author[\v{D}urov\v{c}\'{i}kov\'{a} et al.]{Dominika \v{D}urov\v{c}\'{i}kov\'{a}$^{1}$\thanks{E-mail: dominika.durovcikova@gmail.com}, Harley Katz$^{2}$\thanks{Visitor}, Sarah E. I. Bosman$^{3}$, Frederick B. \newauthor  Davies$^{4}$, Julien Devriendt$^{2}$, and Adrianne Slyz$^{2}$
\\
$^{1}$ New College, University of Oxford, Holywell Street, Oxford OX1 3BN, UK\\
$^{2}$ Sub-department of Astrophysics, University of Oxford, Keble Road, Oxford OX1 3RH, UK\\
$^{3}$ Department of Physics and Astronomy, University College London, Gower Street, London WC1E 6BT, UK\\
$^{4}$ Lawrence Berkeley National Laboratory, CA 94720$-$8139, USA
}

\date{Published 20 February 2020}

\pubyear{2020}

\begin{document}
\label{firstpage}
\pagerange{\pageref{firstpage}--\pageref{lastpage}}\pagenumbering{arabic}
\maketitle

\begin{abstract}
Observations of the early Universe suggest that reionization was complete by $z\sim6$, however, the exact history of this process is still unknown. One method for measuring the evolution of the neutral fraction throughout this epoch is via observing the Ly$\alpha$ damping wings of high-redshift quasars. In order to constrain the neutral fraction from quasar observations, one needs an accurate model of the quasar spectrum around Ly$\alpha$, after the spectrum has been processed by its host galaxy but before it is altered by absorption and damping in the intervening IGM. In this paper, we present a novel machine learning approach, using artificial neural networks, to reconstruct quasar continua around Ly$\alpha$. Our \textsc{QSANNdRA} algorithm improves the error in this reconstruction compared to the state-of-the-art PCA-based model in the literature by 14.2\% on average, and provides an improvement of 6.1\% on average when compared to an extension thereof. In comparison with the extended PCA model, \textsc{QSANNdRA} further achieves an improvement of 22.1\% and 16.8\% when evaluated on low-redshift quasars most similar to the two high-redshift quasars under consideration, ULAS J1120+0641 at $z=7.0851$ and ULAS J1342+0928 at $z=7.5413$, respectively. Using our more accurate reconstructions of these two $z>7$ quasars, we estimate the neutral fraction of the IGM using a homogeneous reionization model and find $\bar{x}_\mathrm{H\Romannum{1}} = 
0.25^{+0.05}_{-0.05}$ at $z=7.0851$ and $\bar{x}_\mathrm{H\Romannum{1}} = 
0.60^{+0.11}_{-0.11}$ at $z=7.5413$. Our results are consistent with the literature and favour a rapid end to reionization.
\end{abstract}




\section{Introduction}

The Epoch of Reionization marked a phase transition in the high-redshift Universe during which neutral hydrogen in the intergalactic medium (IGM) became ionized. The history of reionization and sources responsible are two of the major puzzles of modern cosmology. Recent measurements of the cosmic microwave background (CMB) suggest a substantially neutral IGM at $z \gtrsim 7.5$ \citep{2018arXiv180706209P}.

Quasars constitute one of the most powerful probes of the IGM at high redshifts due to their extremely luminous and non-transient nature (for a full review, refer to \citealt{2016ASSL..423..187M}). The spectra of these distant objects exhibit a damped Ly$\alpha$ emission profile followed by extensive blueward absorption known as the Gunn-Peterson trough \citep{1965ApJ...142.1633G} arising due to damped and resonant absorption by intervening neutral hydrogen. Both of these features have long been recognised as a useful measure of neutral hydrogen density in the intervening gas; however, Ly$\alpha$ absorption saturates at relatively small neutral fractions making the Gunn-Peterson trough suitable for probing the tail-end of reionization only \citep{2006AJ....132..117F}. In contrast, studies of the Ly$\alpha$ damping wing (e.g. \citealt{2011MNRAS.416L..70B,2015MNRAS.454..681K,2018ApJ...864..143D}) provide a wealth of information on the state of the IGM during reionization.

The prerequisite to extracting this information from the damping wings of high-redshift quasars is the knowledge of their intrinsic spectra. For the purpose of this work, we define the term \textit{intrinsic spectrum} as the quasar spectrum after it has been processed by its host galaxy but before it has been affected by the intervening IGM.  If one has a good model for the intrinsic spectrum of a quasar, one can measure the amount of damping needed to transform this intrinsic spectrum to that observed and thus probe the neutral faction in the vicinity of the quasar. 

Fortunately, low-redshift quasars are relatively unaffected by IGM absorption. More than several hundred thousand low-redshift quasars have been observed by the Sloan Digital Sky Survey (SDSS) \citep{2018ApJS..235...42A,2017AJ....154...28B,2013AJ....145...10D,2011AJ....142...72E,2000AJ....120.1579Y}. Strong correlations among the various spectral features in low-redshift quasar spectra have been shown to exist \citep{1992ApJ...398..476F,1992ApJS...80..109B,2004AJ....128.2603Y,2006ApJS..163..110S,2007AJ....134..294S}. Notably, \cite{2017ApJ...840...24E} used these correlations to study reionization through principal component analysis based proximity zone modelling of $z\sim6$ quasars (in \citealt{2018ApJ...864...53E,2019ApJ...881...23E}). Because the portion of the quasar spectra that is significantly redward of Ly$\alpha$ remains relatively unaffected by the intervening neutral IGM, another approach is to develop a model that relates the region of the spectrum redward to Ly$\alpha$ to that which is blueward. This way, the intrinsic spectrum of a quasar can be reconstructed at high-redshift.

\citet{2011Natur.474..616M} and \citet{2018Natur.553..473B} used composite spectra of the most similar low-redshift SDSS quasars to reconstruct the intrinsic spectra of two $z>7$ QSOs.  \citet{2017MNRAS.466.1814G,2019MNRAS.484.5094G} constructed a co-variance matrix to capture the relationships between the Ly$\alpha$, Si\Romannum{4}+O\Romannum{4}], C\Romannum{4} and C\Romannum{3}] emission lines and used these features to reconstruct the intrinsic spectra of the same QSOs. \citet{2018ApJ...864..142D} used a principal component analysis technique to extract the mapping from the spectral features redward of Ly$\alpha$ to the blueward features, and once again applied this to the two $z>7$ QSOs.  However, each of these techniques yields different predictions of the shape of the intrinsic spectra, leading to uncertainties on the high-redshift neutral fraction.

The idea of finding relationships between the intrinsic red side and blue side of QSO spectra is well suited to machine learning.  Much of the physics that governs this relationship is extremely complicated and not well categorised.  For this reason, the mapping between the two regions of a QSO spectrum is non-trivial.  However, the low-redshift SDSS QSO database provides an ideal tool to empirically determine this relationship without making any assumptions on the linearity or the physics involved.  In this paper, we present a novel approach to high-redshift spectra reconstruction termed Quasar Spectra from Artificial Neural Network based predictive Regression Algorithm (\textsc{QSANNdRA}). More specifically, we have implemented an ensemble learning technique by combining 100 artificial neural networks (NNs) into an ensemble called a committee to extract the correlations between the regions of a QSO spectrum redward and blueward of Ly$\alpha$ and thus predict the intrinsic spectrum in a $\sim$100{\AA}-long window around the Ly$\alpha$ peak. Due to the entirely empirical nature of the correlations found in quasar spectra, models like these are promising for extracting the complicated correlations.  More detailed analysis of the features of the model may also reveal some of the underlying physics of the systems.

We present this work as follows. Section~\ref{sec:2} explains how we clean our training data, select an architecture for our neural network, and train our model. In Section~\ref{sec:3}, we apply our model to two of the highest-redshift QSOs known to date, in particular to ULAS~J1120+0641 at $z=7.0851$ \citep{2011Natur.474..616M,2017ApJ...837..146V}, and ULAS~J1342+0928 at $z=7.5413$ \citep{2018Natur.553..473B}, and we use the predictions for these two quasars to constrain the neutral fraction by modelling the damping wing profile. Section~\ref{sec:5} offers a summary and our conclusions.

\section{Methods}\label{sec:2}

To be able to predict the intrinsic spectrum of a high-redshift quasar with strong absorption blueward of Ly$\alpha$, we need to train an algorithm on data for which both the red-side and the blue-side spectra\footnote{For our purposes, \textit{red-side} denotes wavelengths longer than 1290~{\AA}, while \textit{blue-side} denotes wavelengths shorter than 1290~{\AA}.} are observed with minimum or no absorption. Fortunately, the Sloan Digital Sky Survey \citep{2018ApJS..235...42A,2017AJ....154...28B,2013AJ....145...10D,2011AJ....142...72E,2000AJ....120.1579Y} has collected data of several hundred thousand low-redshift quasars that are suitable for this purpose. This section explains how we select and clean the low-redshift SDSS data to build our model, while measuring its applicability to high-redshift quasars.

\subsection{Data cleaning and training set compilation}\label{sec:2.1}

In this section, we explain the data cleaning 
procedure and define the criteria used to select the 
low-redshift data, thus compiling the preliminary training 
set for our model.

All low-redshift quasar spectra come from 
the Extended Baryon Oscillation Spectroscopic Survey 
(eBOSS) \citep{2016AJ....151...44D} of the Sloan 
Digital Sky Survey \Romannum{4} (SDSS-\Romannum{4}) 
\citep{2018ApJS..235...42A,2017AJ....154...28B} and 
its earlier phases 
\citep{2013AJ....145...10D,2011AJ....142...72E,2000AJ....120.1579Y}. The primary training data selection 
was performed based on the fourteenth data release 
version of the SDSS Quasar Catalog 
\citep{2018A&A...613A..51P} using criteria that were 
mostly inspired by \citet{2018ApJ...864..143D}.

In order to define the training set, we firstly identified all available quasars with \texttt{Z{\_}PIPE} redshifts from 2.09 to 2.51. This redshift range enables us to capture the entire Ly$\alpha$ peak as well as other significant features of the spectra, such as the C\Romannum{4} and the Mg\Romannum{2} emission lines. We rejected all QSOs with highly uncertain redshifts (\texttt{ZWARNING}$\neq$0) and all broad-absorption-line quasars (BALs, \texttt{{BI{\_}CIV}}$\neq$0). These cuts reduced the data set to 101,739 QSOs. We then performed a signal-to-noise ratio cut \texttt{SN{\_}MEDIAN{\_}ALL}$>$7.0 which significantly reduced the number of QSOs to 19,054. In Appendix~\ref{SN_app} we explore the effect that varying the S/N threshold value has on the overall performance of our model.

\begin{figure*}
    \centering
    \includegraphics[width=\linewidth]{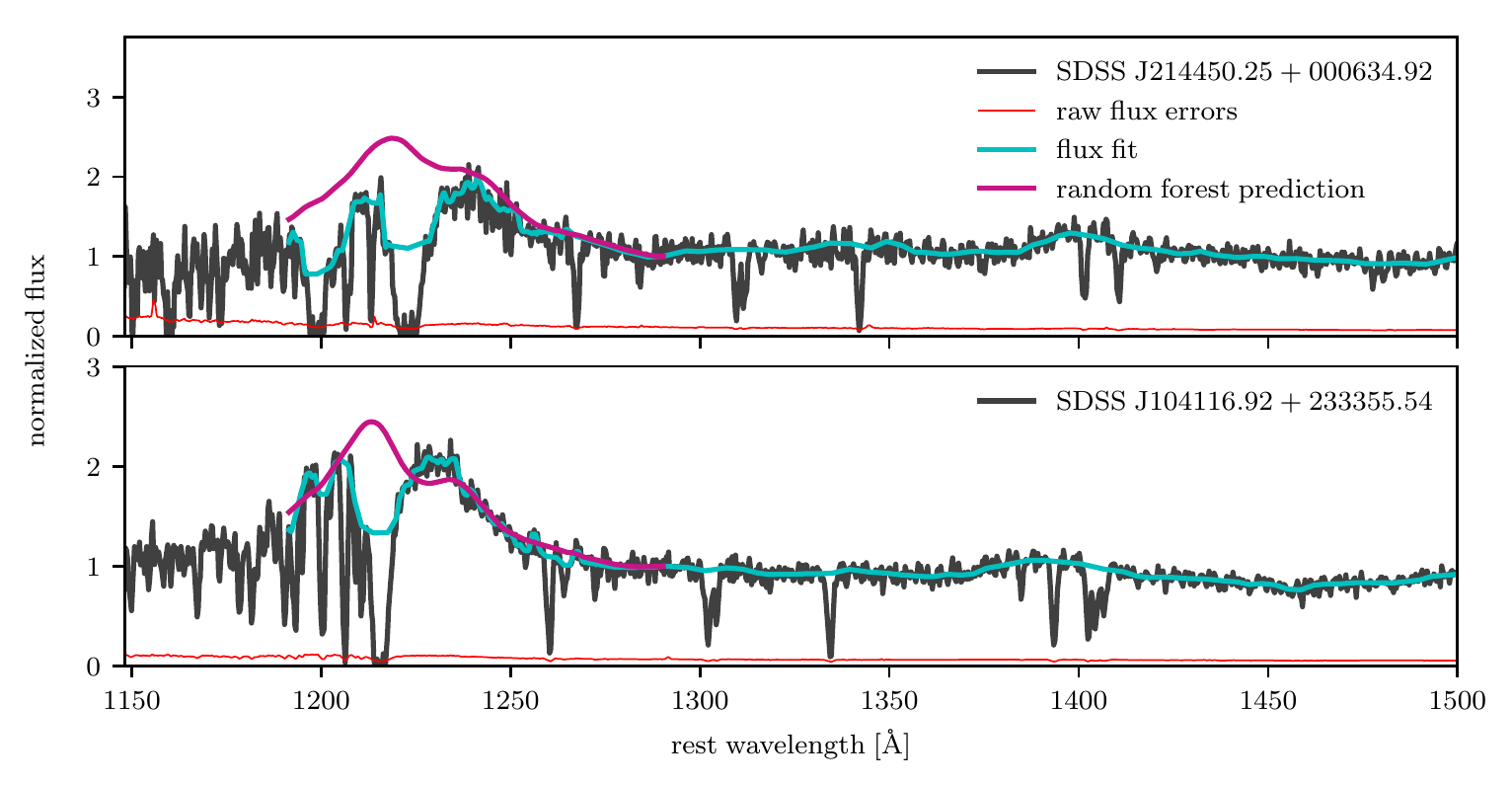}
    \caption{Two examples of spectra with strong 
    absorption features blueward of 1290{\AA} that 
    were rejected based on preliminary random forest 
    predictions. The raw smoothed data for each 
    quasar are shown in gray, the flux fit is shown 
    in cyan and our blue-side predictions are shown 
    in magenta. We also show the flux errors on the 
    raw data as a thin red curve. Note that the y-axis is normalized with respect to each quasar's 
    fitted flux at 1290{\AA}. }
    \label{fig:rejected}
\end{figure*}

For each spectrum, we masked out all sky lines listed in Table~30 of \citet{2002AJ....123..485S} as well as all pixels that were flagged as highly uncertain. All remaining spectra were subsequently smoothed. A detailed account as well as a visual demonstration of the smoothing procedure is provided in Appendix~\ref{sec:smooth_app}, and hence we only provide an outline below for the sake of brevity. We first computed a running median with a bin size of 50 data points to capture the main continuum and emission features in the spectrum. We then performed a peak-finding procedure using the SciPy Python library \citep{scipy} above the aforementioned running median border and interpolated the peaks to construct an upper envelope of the spectrum. This envelope was then subtracted from the spectrum. We then applied the RANSAC regressor algorithm \citep{fischler1981random} from the Scikit-Learn Python package \citep{2012arXiv1201.0490P} on the residuals, thus rejecting most absorption features in the spectrum. The data points that were flagged as inliers by RANSAC were interpolated and smoothed by computing a running median with a bin size of 20, thus creating the final smooth flux fit of each spectrum.

The smoothed spectra were used to perform the final set of cuts. First, the observed wavelengths, $\lambda_\mathrm{obs}$, were calibrated to rest wavelengths, $\lambda_\mathrm{rest}$, according to
\begin{equation}
	\lambda_\mathrm{rest} = \frac{\lambda_\mathrm{obs}}{1+z},
\end{equation}
where $z$ is the object's SDSS redshift coming from the broad UV emission lines. In Appendix~\ref{z_app} we explore the systematic errors potentially associated with this redshift calibration. Next, we normalised the spectra such that all fluxes at 1290{\AA} were equal to unity, and then rejected all quasars whose fitted fluxes fell below 0.5 blueward of 1280{\AA} or below 0.1 redward of 1280{\AA}. This was done in order to reject quasars with strong associated absorption or poor signal-to-noise ratio redward of Ly$\alpha$, respectively. It should also be noted that this normalisation also removes the sensitivity to the Baldwin effect \citep{1977ApJ...214..679B} as well as the correlation between the quasar brightness and emission line shifts \citep{2003ApJ...586...52S,2011AJ....141..167R}. The significance of this step is discussed in Section~\ref{sec:5}. The remaining data set consists of 17,007 quasars, whose fluxes were interpolated at 3,862 different wavelengths between 1191.5{\AA} and 2900.0{\AA} (spaced uniformly in log space).

\subsection{Refining the training set with Random Forests}\label{sec:2.2}

Visual inspection of the data revealed that our training set still contained a few spectra with strong absorption features blueward of 1290{\AA}. This section describes a random forest (RF) based procedure we used to further reject these quasars from the training set.

Random forest regression is an ensemble learning algorithm, which combines multiple decision trees to form a statistical prediction of the output value, or the blue-side flux in our case. Within the random forest, each decision tree makes a flux estimation after a series of queries on the red-side spectral properties. The overall predicted flux is then determined by taking the average of predicted flux values by all the trees in the forest.

The power of random forests results from the way the trees are grown in the training phase. Growing each individual tree is done by a random selection of spectral features and training spectra (i.e. bagging or bootstrapping) based on which the tree learns to make the prediction. The training is performed by feeding example, or training, spectral data into the forest, where each tree gradually learns to map the input (red-side) values to the output (blue-side) ones by comparing its own prediction to what the output is supposed to be.

The motivation behind choosing a RF regressor \citep{Breiman2001}, as implemented in the Scikit-Learn Python package \citep{2012arXiv1201.0490P}, is that RFs are rather easily trained and are useful in confirming strong correlations between the red-side and the blue-side spectral features. Because of these correlations, it is likely that the RF would be unable to predict uncorrelated absorption features, which makes its predictions a suitable tool to detect the remaining quasars with strong absorption in the training set. In other words, the RF will have a large prediction error on the few remaining QSOs in our data set that have strong absorption features around Ly$\alpha$ and thus this high prediction error is indicative of an outlier in our data set. Furthermore, because each tree learns the red-side to blue-side mapping based on a different set of criteria and training spectra, we can prevent overfitting even when using decision trees that are arbitrarily deep. The downsides are that training large RFs is extremely memory-intensive and that they may not generalise as well as other machine learning algorithms (such as a neural network).

To train a random forest, we split the training set into train and test subsets (80:20 ratio). We standardised the fluxes to have a mean of 0 and a variance of 1 for each wavelength across all objects, upon which we performed principal component analysis (PCA) using the Scikit-Learn Python package \citep{2012arXiv1201.0490P} to reduce the feature space and complexity of the random forest. Standardisation was performed in order to make the different quasars as well as the different features in each spectrum comparable, which improves the PCA search for vectors of maximum variance. We chose the number of principal components such that they capture 99\% of variance in the data both on the red side and the blue side (we will henceforth refer to this number as the \textit{explained variance ratio}). After this transformation, the data were fed into a RF with 100 decision trees. 

Inspection of the preliminary RF predictions revealed that, as expected, the RF predicted large errors for the QSOs in the data set that exhibited strong absorption features.  We took advantage of this feature to remove most of the remaining quasars with strong absorption from our training set as follows. We defined the relative prediction error, $\epsilon$, to be the relative absolute error of the prediction against the smoothed flux value at a particular wavelength, or
\begin{equation}\label{eq:epsilon}
	\epsilon = \frac{\mid{F_\mathrm{pred}-F_\mathrm{smooth}}\mid}{F_\mathrm{smooth}},
\end{equation}
where $F_\mathrm{pred}$ is the flux predicted by the RF and $F_\mathrm{smooth}$ is the smoothed flux. Based on the error for each predicted data point in the test set, we rejected all data points whose relative prediction error was greater than $\bar{\epsilon} + 3\sigma_\mathrm{\epsilon}$, where $\bar{\epsilon}$ is the mean error and $\sigma_\mathrm{\epsilon}$ is the standard deviation of the error across all objects for a particular wavelength. This procedure was repeated 10 times, each time training the random forest on 9 subsets of the whole training set (henceforth termed \textit{folds}) and rejecting data points in the 10th one. This way, 1.1\% of all data points on the blue side were rejected in the whole training set, altogether rejecting 3,304 objects. This left us with a cleaner training set of 13,703 spectra. Figure~\ref{fig:rejected} displays two typical examples of spectra that were rejected in this process.

\subsection{Construction of \textsc{QSANNdRA}}\label{sec:2.3}

\begin{figure}
    \centering
    \includegraphics[width=\columnwidth]{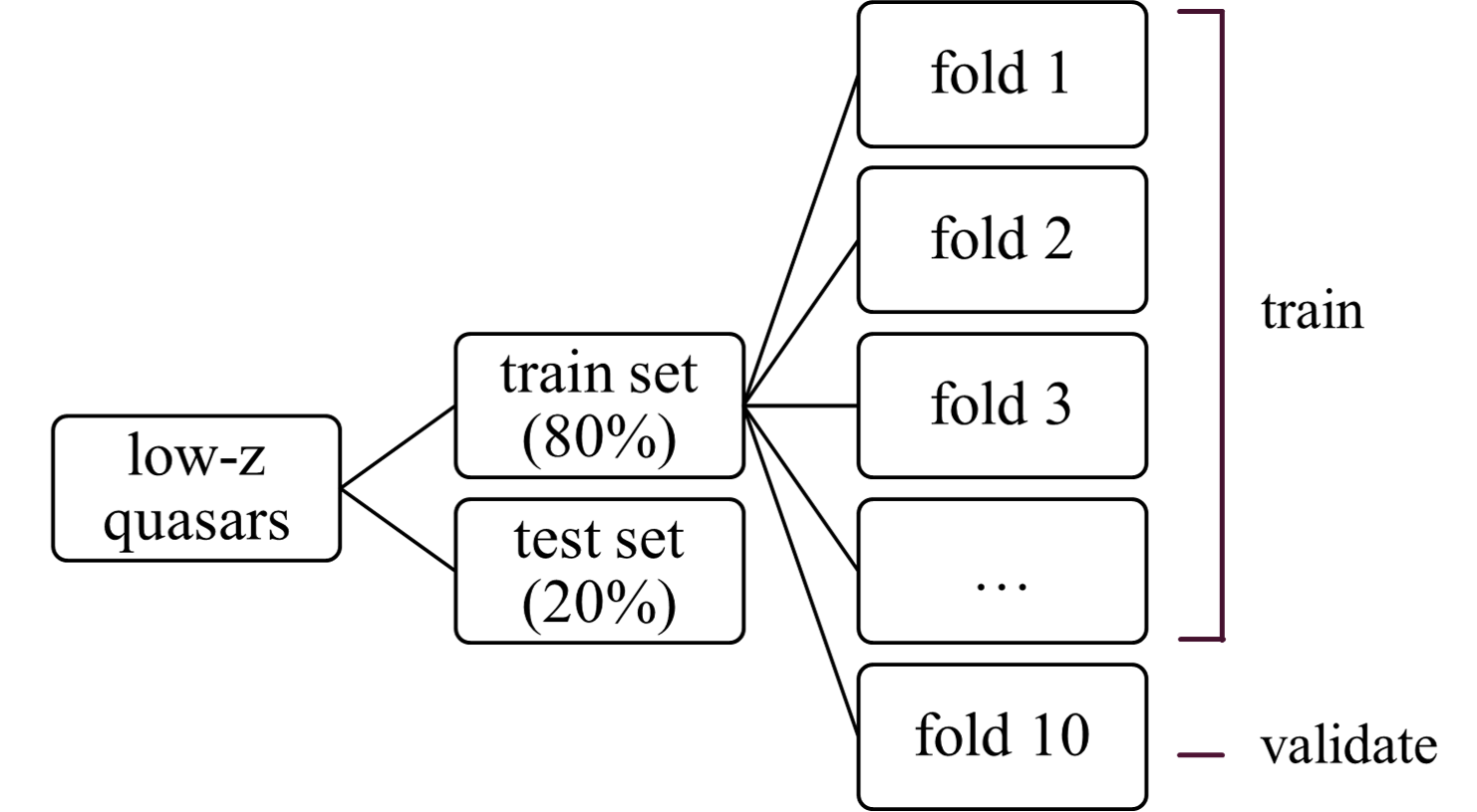}
    \caption{Infographic showing how 
    the low-redshift data were divided 
    for 10-fold crossvalidation. The 
    whole training set was divided into 
    a train and a test set (80\% and 
    20\% of the quasars, 
    respectively), and the train set 
    was further subdivided into 10 
    folds. During crossvalidation, the 
    neural network was trained 10 
    times, each time training on 9 
    folds and validating its 
    performance on the remaining fold 
    to assess generalizability.}
    \label{fig:k-fold}
\end{figure}

\begin{figure}
    \centering
    \includegraphics[width=\columnwidth]{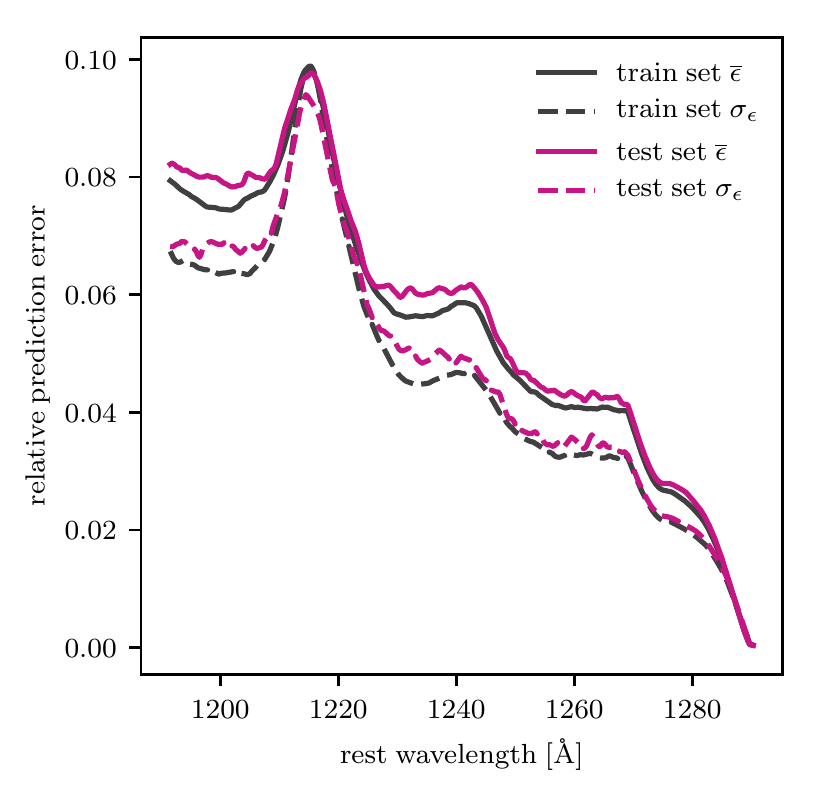}
    \caption{Mean (solid curve) and standard 
    deviation (dashed curve) of the relative 
    prediction error as defined in 
    equation~\ref{eq:epsilon} for the neural network 
    blue-side predictions as a function of 
    wavelength. The set of gray 
    curves shows the NN's performance on the train 
    set, while the set of magenta curves shows its 
    performance on the test set (i.e. previously 
    unseen data). }
    \label{fig:NN_er}
\end{figure}

This section outlines the implementation of a feed-forward neural network (NN) on our training set and further describes the construction and training of our predictive model called \textsc{QSANNdRA}.

\begin{figure*}
    \centering
    \includegraphics[width=\linewidth,trim={0cm 0.7cm 0cm 0cm},clip]{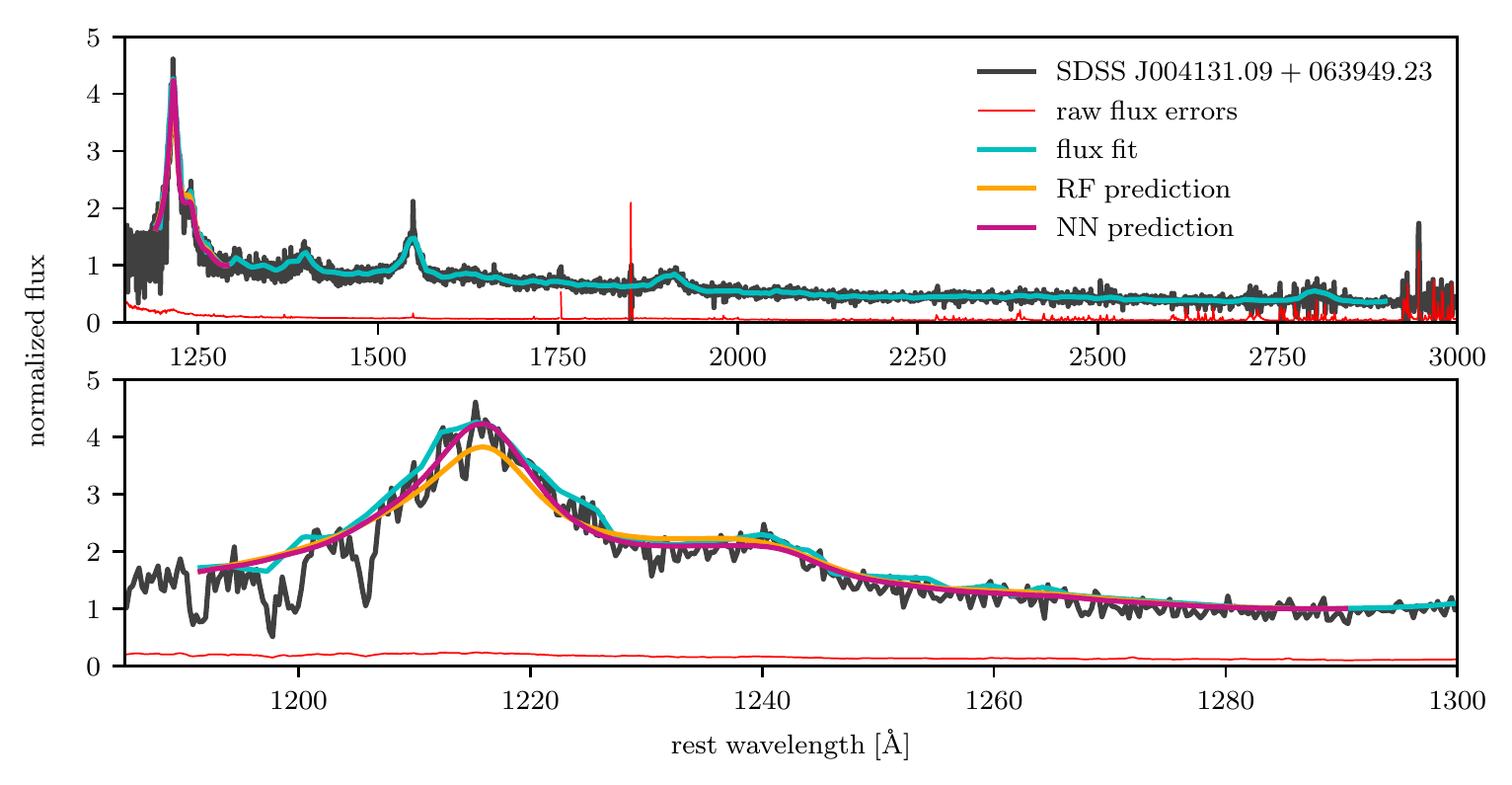}
    \includegraphics[width=\linewidth,trim={0cm 0cm 0cm 0cm},clip]{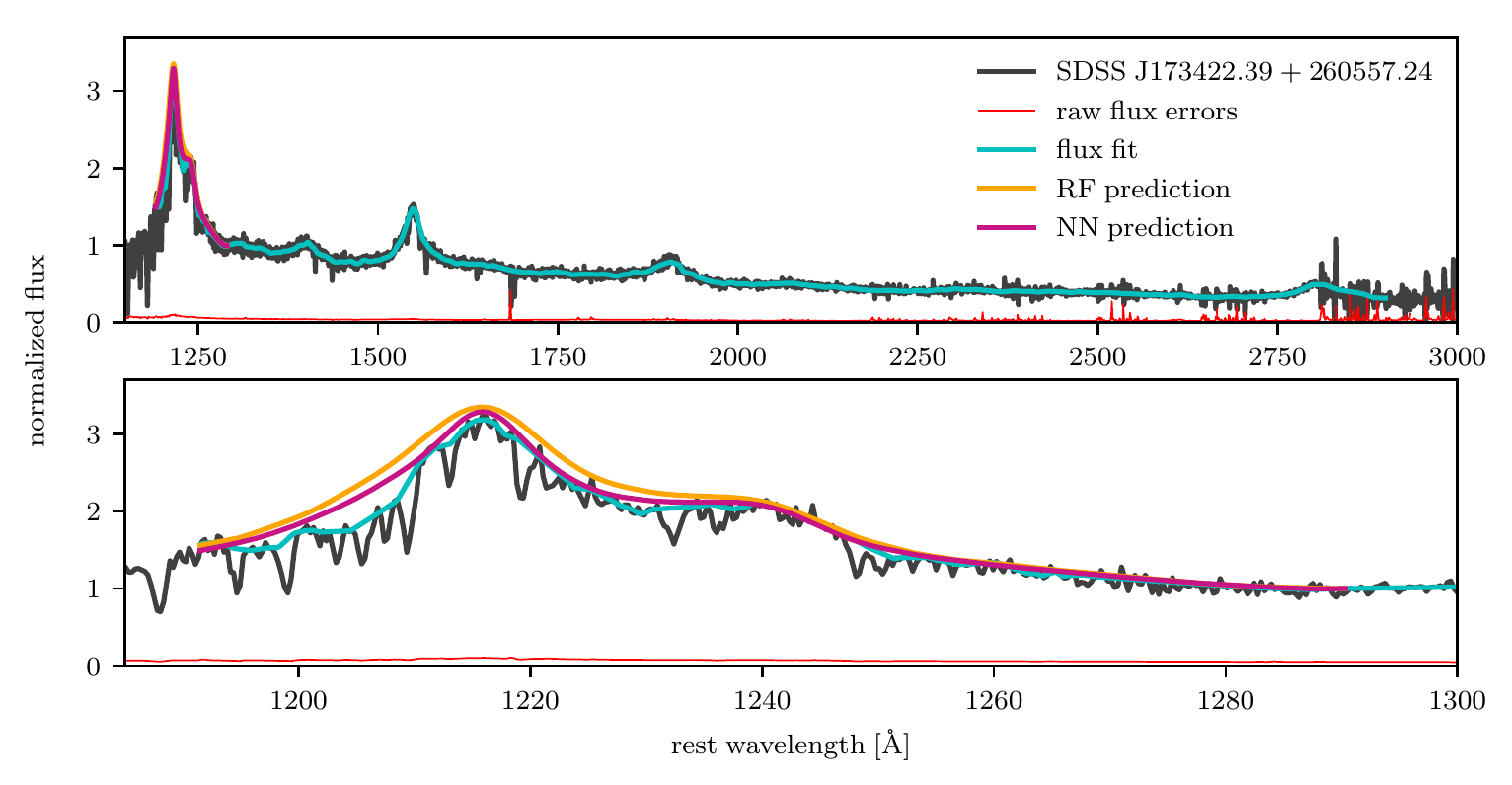}
    \caption{Two examples of low-redshift quasar spectra drawn from the test set comparing the RF (orange) and NN (magenta) predictions. The raw data are depicted in gray, the errors on the raw data are shown as a thin red curve, and the flux fit is shown in cyan. The top panel for each quasar shows the full spectrum, while the bottom panel offers a close-up view of the predictions. Note 
    that the y-axis is normalised with respect to 
    each quasar's fitted flux at 1290{\AA}.}
    \label{fig:NN_RF_examp_1}
\end{figure*}

We implemented an artificial neural network (NN) in order to better capture the correlations between the different spectral features of QSOs. NNs are stacks of interconnected layers of computational units called neurons, where each neuron is assigned a set of weights, $w_\mathrm{i}$, and a bias, $b$. It multiplies all its inputs, $x_\mathrm{i}$, by the corresponding weights and adds the bias before passing the result on to the following neurons in subsequent layer. Each layer is then assigned an activation function $f$, so that the output of a neuron in that layer, $y$, becomes
\begin{equation}
	y = f \Bigg{(}b + \sum_{i} w_ix_i\Bigg{)} = f(b + \pmb{w}^\mathrm{T} \pmb{x}).
\end{equation}
The power of even the simplest NNs is that they allow for modelling of non-linear relationships between the input and the output. This is due to the fact that the information from the different neurons is combined in a non-linear fashion as it propagates through the layers of the network. In an untrained network, the weights and biases on each neuron are typically initiated randomly. Then, by iteratively passing training inputs (i.e. red-side spectra) through the network, comparing its outputs (i.e. blue-side spectra) to what the training outputs should be and updating the weights and biases along the gradient of the loss function\footnote{The loss function defines the discrepancy between the predicted flux value and the real flux value.} with respect to that weight, learning of the ANN is achieved. More detailed information on NNs can be found in \citet{geron2017hands}.

We implemented a fully connected feed-forward NN using the Keras Python package \citep{chollet2015keras}, which means that each neuron was connected to all neurons in the preceding and following layer, and that information was propagated in only one direction from the input to the output. Since we operate with 63 and 36 principal components on the red side and the blue side, respectively, the number of neurons in the input and output layers was fixed at 63 and 36. To further define its architecture and hyperparameters, i.e. number of layers, number of neurons in each layer, activation function, number of training epochs\footnote{One epoch is defined as passing the full training data set through the network once.} and the batch size\footnote{The batch size is the number of training examples that are passed through the network before the weights get updated.}, we performed an extensive grid search over 100 different networks and training settings. For each combination, 10-fold cross-validation (see Figure~\ref{fig:k-fold}) was performed to ensure generalisability, i.e. a comparable performance when applied to previously unseen data. For cross-validation, the training subset was further divided to 10 folds, and the NN training was repeated 10 times, each time training on 9 folds and computing the mean-absolute-error (i.e. the loss function) of the prediction on the 10th fold. The mean error and its standard deviation was reported for each NN, and these scores were then compared to choose the NN with the best performance. To further fine-tune the training parameters, i.e. number of epochs and the batch size, a further, smaller grid search over these parameters was performed for the best-performing NN architecture.

The NN with the best performance was found to have the following architecture: 63-40-40-36\footnote{Each number stands for the number of neurons in the corresponding layer.} with the `elu' activation function\footnote{The `elu' function is defined as:
\begin{equation*}
	\begin{split}f(x) = \begin{Bmatrix} x & x > 0 \\
 \alpha(e^x - 1) & x \leq 0 \end{Bmatrix}\end{split},
\end{equation*}
where $-\alpha$ defines the horizontal asymptote at negative infinity.}, and was trained for 80 epochs using a batch size of 500 QSOs. When applied to the test set, the mean relative prediction error was 5.7\% per data point. Figure~\ref{fig:NN_er} displays how $\bar{\epsilon}$ (solid curves) and  $\sigma_\mathrm{\epsilon}$ (dashed curves) vary across the predicted wavelength range both for the train set (gray set of curves) and the test set (magenta set of curves).  Note how there are distinct features in the error of our model as a function of rest~frame wavelength.  In general, the redder the wavelength, the less error we predict for our model.  This is simply due to the fact that the closer the wavelength is to the known data, the more reliable the extrapolation is.  The bumps in the error are due to the presence of emission lines in the QSO spectra.  Most notably, the three bumps that we see are due to Ly$\alpha$, NV at 1240{\AA}, and SiII at 1260{\AA}, from blue to red respectively. The strengths and velocity shifts of these emission lines are more difficult to predict than the underlying continuum and hence the error is enhanced around their wavelengths.  Figure~\ref{fig:NN_RF_examp_1} shows two examples of low-redshift quasars drawn from the test set, where we have used our NN and the earlier trained RF to predict the QSO continuum around Ly$\alpha$.  In both examples, our predictive model performs very well.

\begin{figure}
    \centering
    \includegraphics[width=\linewidth]{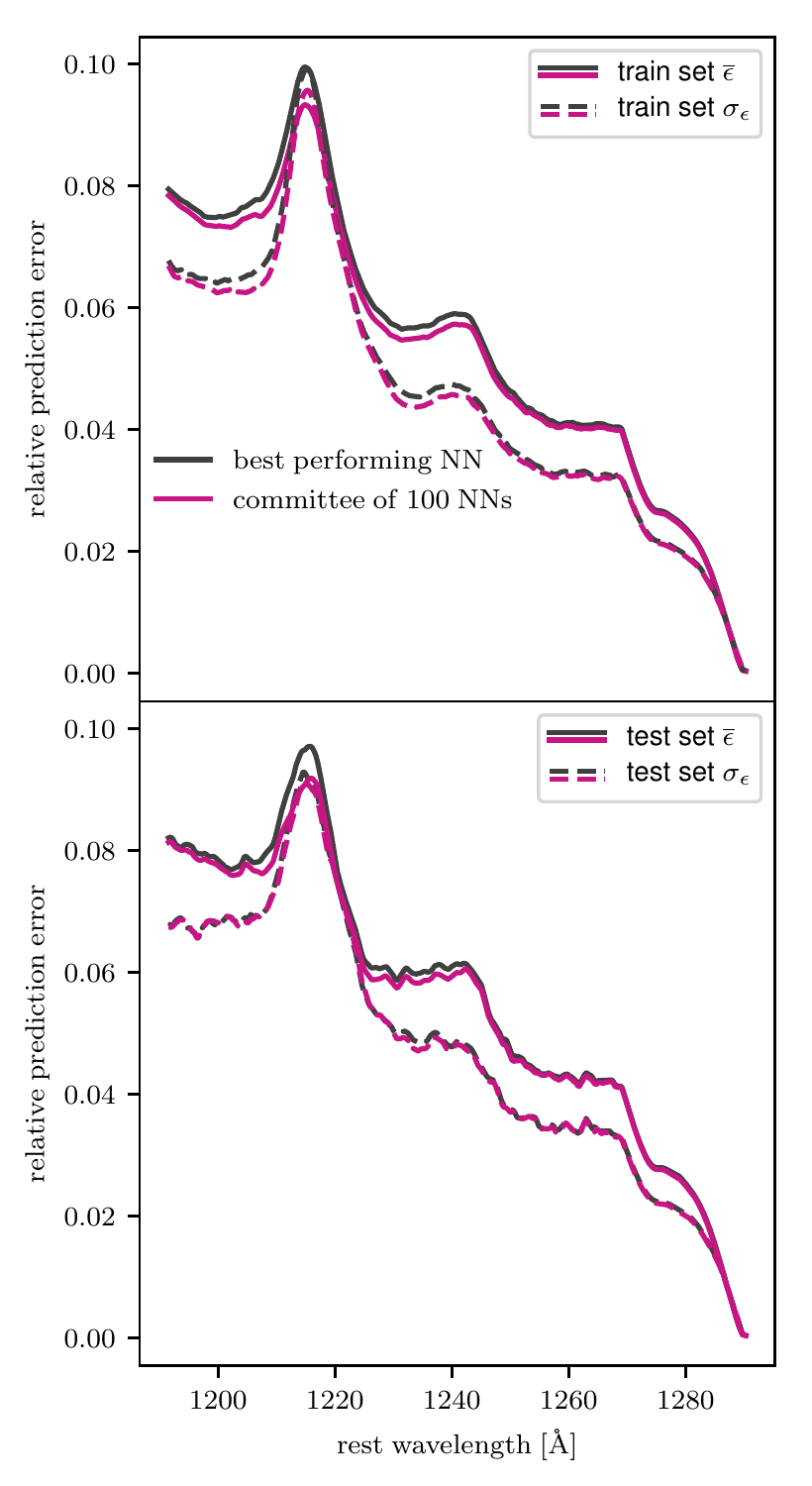}
    \caption{A comparison of the relative prediction error statistics as defined by equation~\ref{eq:epsilon} for the committee of 100 networks (magenta) to the best performing neural network (gray) within the committee for the train set (top) and the test set (bottom). The solid curves show the mean relative prediction error, and the dashed curves represent its standard deviation. The ${\sim}$5\% decrease in mean relative error around the Ly$\alpha$ peak in both train and test sets confirms the improvement in both accuracy and robustness upon implementing the ensemble technique.}
    \label{fig:best_to_100}
\end{figure}

To improve the performance and increase the robustness of our method, we trained 99 more neural networks with the same architecture and hyperparameters. The weights and biases of each of the neurons were initiated using a different random seed. This way, we created a committee of NNs, which falls under ensemble learning techniques (see \citealt{dietterich2000ensemble}). The idea behind ensemble learning is that if the errors made by individual predictors are uncorrelated, they will cancel out with each other when averaging the predictions from multiple algorithms. Furthermore, it is possible for a neural network to get trapped in a local minimum during training and thus never reach the optimal solution. Initiating the weights and biases in each neural network differently provides a different starting point for each training, which can result in different networks converging to different local minima. Averaging these locally optimal predictions has the potential to come nearer to the globally optimal prediction and thus decrease the overall prediction uncertainty.

We trained each network separately and evaluated the performance of each of them on the test set of the low-redshift SDSS data. The inverse of the achieved mean relative prediction error $\bar{\epsilon}$ then determined the weight assigned to each of the individual neural networks in the ensemble. Subsequently, the weighted predictions were averaged to produce the overall prediction of the committee, whose performance was once again evaluated on the low-redshift test set. This method improved the mean relative prediction error to 5.5\% per data point (as compared to 5.7\% for the individual NN described previously).

\begin{figure}
    \centering
    \includegraphics[width=\columnwidth]{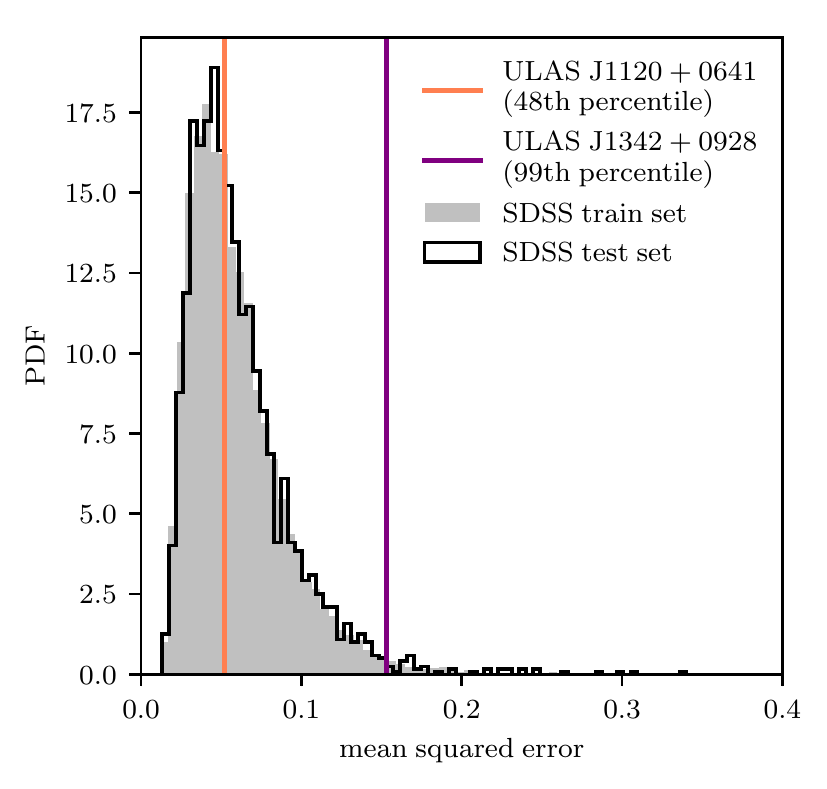}
    \caption{A histogram of autoencoder-produced mean squared error for our low-redshift dataset as compared to the mean squared error for ULAS J1120+0641 and ULAS J1342+0928. The distribution of the train set is displayed in gray, the distribution of the test set is displayed in black, while the orange and purple lines indicate the position of ULAS J1120+0641 (48th percentile) and ULAS J1342+0928 (99th percentile), respectively, within that distribution. Note that the percentile here defines how well the spectral features of the two high-redshift quasars are represented in the training sample of low-redshift SDSS quasars (i.e. the larger the percentile, to more outlying the quasar with respect to the low-redshift dataset). }
    \label{fig:hist}
\end{figure}

It should be emphasised that the power of this approach lies in the potential of the ensemble outperforming even the best algorithm within the ensemble. In fact, as shown in Figure~\ref{fig:best_to_100}, the committee achieved a ${\sim}$5\% improvement on the mean relative prediction error around the Ly$\alpha$ peak as compared to the best performing NN in our ensemble for both the train and test set predictions. Moreover, we observe that the ensemble achieves a mean relative prediction uncertainty spanning from $\sim$5\% to $\sim$9\% in the wavelength range $\sim$1210{\AA} to $\sim$1250{\AA} most relevant to damping wing modelling, with the standard deviation spanning from $\sim$4\% to $\sim$7.5\%. The difference between the performance of the committee on the train and test sets is marginal, which suggests a strong generalizability of our model to new data. The validation stage of the ensemble's performance thus confirms an improved accuracy and robustness in using the committee of networks as opposed to just one individual prediction.

\subsection{Applicability of \textsc{QSANNdRA} to \texorpdfstring{$z>7$}{z<7} QSOs}\label{sec:2.4}

\begin{figure*}
    \centering
    \includegraphics[width=\linewidth]{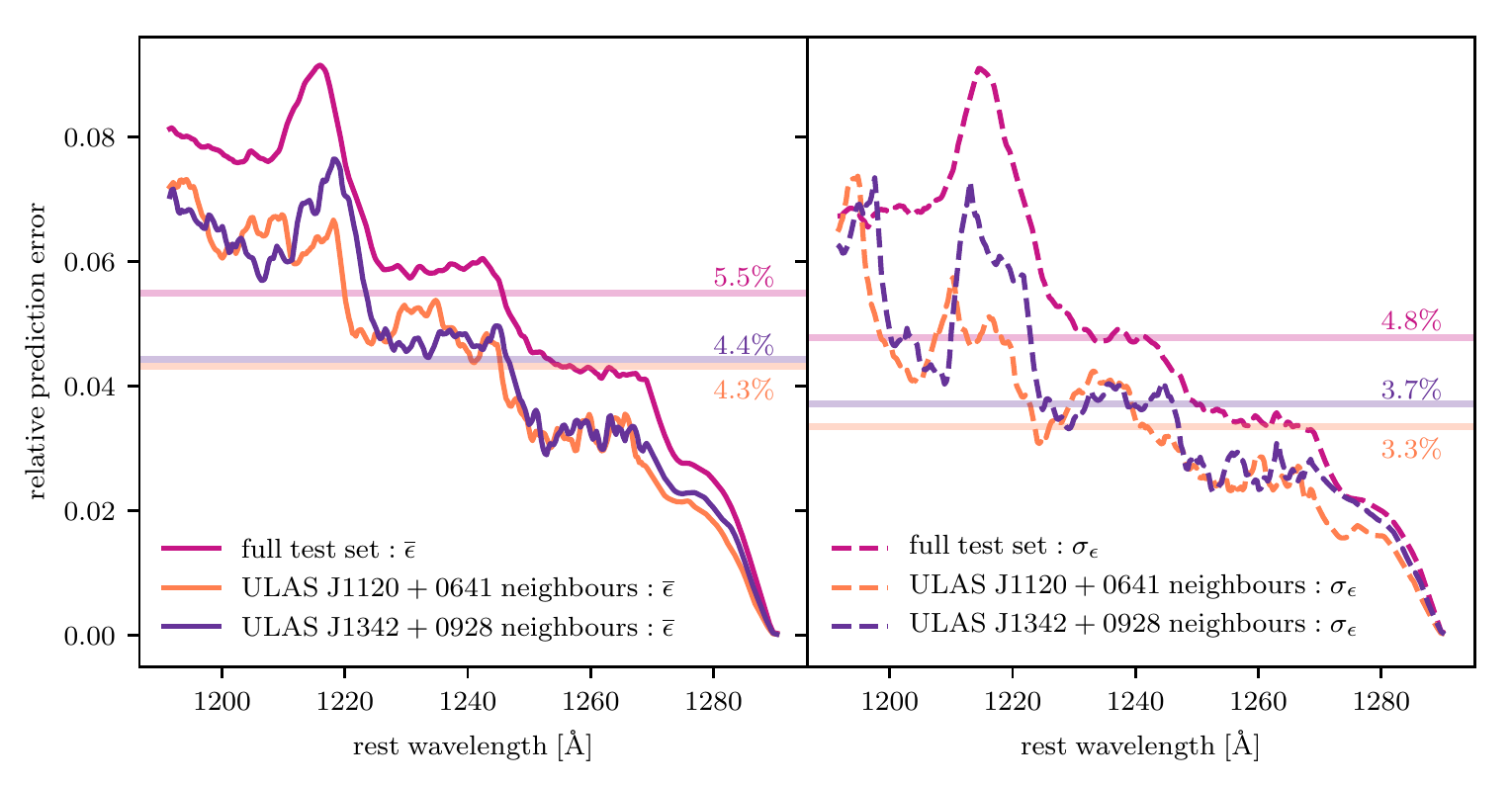}
    \caption{Wavelength-dependent distribution of $\bar{\epsilon}$ (left) and $\sigma_\mathrm{\epsilon}$ (right) for \textsc{QSANNdRA}'s full test set performance (magenta) as compared to its performance on 100 nearest neighbour low-redshift quasars from the test set for ULAS J1120+0641 (orange) and ULAS J1342+0928 (purple). Note that even though the $z=7.5$ QSO has been shown to be on the tail-end of the low-redshift quasar distribution, \textsc{QSANNdRA} actually performs better on low-redshift quasars which are similar to it.}
    \label{fig:neighbours}
\end{figure*}

Although we have developed a robust algorithm for predicting the intrinsic QSO spectra of low-redshift SDSS QSOs, it is important to determine the applicability of this model to the high-redshift quasars that we aim to use to constrain the neutral gas fraction during the Epoch of Reionization. In particular, we aim to apply our model to the combined VLT/FORS + Gemini/GNIRS spectrum of ULAS J1120+0641 \citep{2011Natur.474..616M} and the Magellan/FIRE + Gemini/GNIRS spectrum of ULAS J1342+0928 \citep{2018Natur.553..473B}. If there is a fundamental difference between the high-redshift quasars and the SDSS QSOs, even though the trained NNs are meant to be generalizable, their predictive power on such different systems deserves to be questioned.  For this reason, in this subsection, we provide a method to determine both how similar the two $z>7$ QSOs are to the SDSS quasars as well as the performance of \textsc{QSANNdRA} on the low-redshift QSOs that are most similar to those at $z>7$.

To quantify how unusual the two high-redshift quasars are, we trained an autoencoder on the red-side PCA components of the low-redshift SDSS data. An autoencoder is a neural network with two components, an encoder and a decoder. The first compresses the input data while the second reconstructs it. It essentially acts as an identity function. Training the autoencoder on the SDSS QSOs causes the network to pick out the spectral features that are most represented in our low-redshift training set, which then form the basis for reconstruction. By measuring the error between the input and the reconstructed output coefficients, one can determine how effective the compression of the data is for all the SDSS quasars, and hence define an error distribution that is characteristic of our low-redshift sample. Subsequently, by running the trained autoencoder on the red-side PCA coefficients of the two high-redshift quasars and measuring the error, we are able to quantify how well represented the red sides of the high-redshift quasars are in our training dataset and hence determine the extent of their outlying nature.

We thus trained a 4-layer autoencoder whose input and output layers both had 63 neurons (corresponding to 63 red-side PCA components) and the middle two layers were composed of 30 neurons. The activation function on all neurons was set to `elu'. We trained the autoencoder for 100 epochs in batch sizes of 500 while optimizing for the mean squared error between the input and output coefficients. We then applied the trained autoencoder on the test set of low-redshift quasars. We further calculated the resulting mean squared error across all red-side coefficients for each quasar both in the train set and the test set and composed a probability density function for this error for the low-redshift dataset.

Applying the trained autoencoder to ULAS~J1120+0641 and ULAS~J1342+0928 and calculating their corresponding mean squared errors revealed that they fall onto the 48th percentile and the 99th percentile in our low-redshift distribution (see figure~\ref{fig:hist}). While ULAS~J1120+0641 seems to be well represented in our low-redshift dataset, ULAS~J1342+0928 falls onto the tail-end of our distribution. These findings are comparable to those of \cite{2018ApJ...864..142D}, who used a mixture of multivariate Gaussians to determine a percentile of 15$\%$ and 1.5$\%$ for the two high-redshift QSOs with respect to their low-redshift dataset (values equivalent to our percentiles of 85$\%$ and 98.5$\%$). While these agree that the $z=7.5$ QSO is a 2$\sigma$ outlier, our training set seems to be more representative of the $z=7.1$ QSO than that of \citep{2018ApJ...864..142D}. Especially in light of the outlying nature of ULAS J1342+0928, it is crucial to assess the performance of our model on its nearest-neighbour low-redshift quasars.

Motivated by Figure~\ref{fig:hist}, we further evaluated \textsc{QSANNdRA}'s performance on 100 nearest-neighbour QSOs of ULAS~J1120+0641 and ULAS~J1342+0928 (i.e. the QSOs that are most similar to the high-redshift quasars). We chose the 100 quasars by computing the Euclidean distance between the red-side PCA coefficients of each of the high-redshift quasars and the red-side PCA coefficients of the low-redshift quasars in our test set only. Only searching through the test set of SDSS quasars, which the model did not see during the training stage, guarantees a generalizable performance. The resulting $\bar{\epsilon}$ and $\sigma_\mathrm{\epsilon}$ distributions across the blue-side wavelengths are shown in Figure~\ref{fig:neighbours}, where the performance on 100 nearest-neighbours of ULAS~J1120+0641 and ULAS~J1342+0928 are shown in orange and purple, respectively, and the full test set performance is shown in magenta for comparison. \textsc{QSANNdRA} performs better on SDSS quasars that are similar to the high-redshift ones (the full test set $\bar{\epsilon}$ is 5.5\% on average as compared to 4.3\% and 4.4\% on average for the nearest neighbours of ULAS~J1120+0641 and ULAS~J1342+0928, respectively; the full test set $\sigma_\mathrm{\epsilon}$ is 4.8\% on average as compared to 3.3\% and 3.7\% on average for the nearest neighbours of ULAS~J1120+0641 and ULAS~J1342+0928, respectively) than on the full test set, which hints at a greater reliability of our predictions for these two high-redshift quasars and confirms that \textsc{QSANNdRA} should be able to tackle the outlying spectral features of ULAS~J1342+0928.

\subsection{\textsc{QSANNdRA} compared to existing models}

\begin{figure*}
    \centering
    \includegraphics[width=\linewidth]{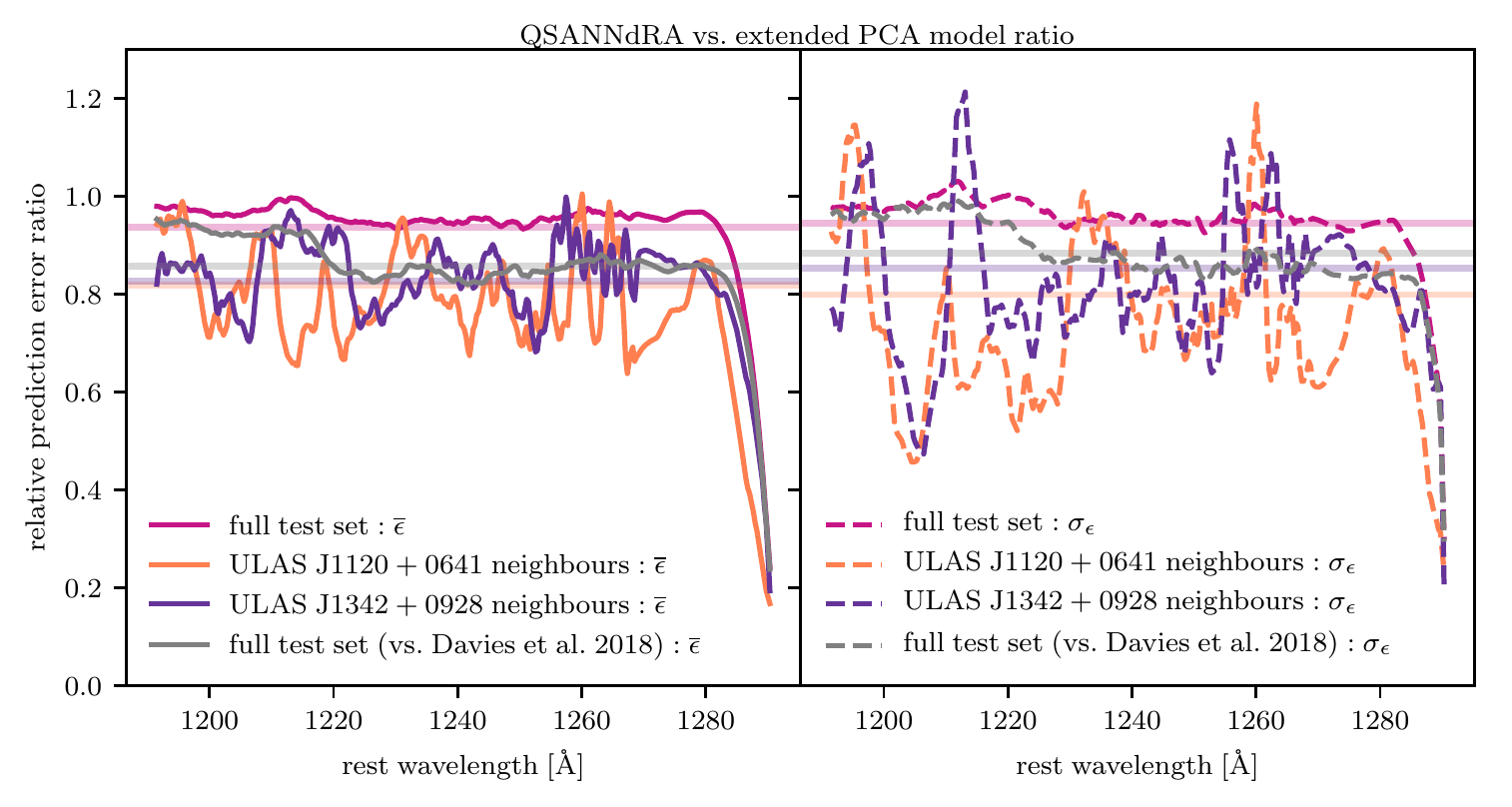}
    \caption{Wavelength-dependent distribution of the ratio of \textsc{QSANNdRA}'s performance to the original PCA model's performance reported by \protect\cite{2018ApJ...864..142D} as well as the extended PCA model's performance in terms of $\bar{\epsilon}$ (left) and $\sigma_\mathrm{\epsilon}$ (right). The full test set performance improvement as compared to the published model \citep{2018ApJ...864..142D} is shown in gray. When compared to the extended PCA model, the full test set performance ratio (magenta) is compared to the performance ratios on the 100 nearest low-redshift quasar neighbours of ULAS~J1120+0641 (orange) and ULAS~J1342+0928 (purple) from the test set.}
    \label{fig:davies_ratio}
\end{figure*}

Before proceeding, it is important to understand how the predictive power of the \textsc{QSANNdRA} model compares to other models available in the literature. To better quantify the performance of \textsc{QSANNdRA} to existing models, we implemented the state-of-the-art PCA-based model from \cite{2018ApJ...864..142D} and an extension thereof (explained below) on our cleaned training data set to directly compare the results. Note that there are some differences in our implementation compared to the model published by \cite{2018ApJ...864..142D} which makes it more comparable to the traditional PCA-based techniques \citep{2005ApJ...618..592S,2011A&A...530A..50P}. Our cleaning and smoothing procedures are slightly different, we do not use nearest-neighbour stacks to compute the PCA basis vectors, and we do not fit for redshifts simultaneously. In contrast to the original paper \citep{2018ApJ...864..142D}, our training data set relies on a different version of the SDSS database, and we also split our low-redshift SDSS data set into train and test sets to ensure that we are making a fair comparison between the models' performances on previously unseen data.

We implemented the model by \cite{2018ApJ...864..142D} with 10 red-side and 6 blue-side PCA components, and solved for a linear mapping between the red-side and the blue-side coefficients using a least-squares solver. We then computed the blue-side coefficients for all quasars in the test set and $\bar{\epsilon}$ and $\sigma_\mathrm{\epsilon}$ for each blue-side wavelength in each case. 

As an extension to the original model \citep{2018ApJ...864..142D}, we adjusted the number of principal components in this model to be the same as in our model to work with an equivalent amount of information (63 on the red side and 36 on the blue side). We will henceforth refer to this model as the extended PCA model. We repeated the same procedure as described above to assess the improvement of \textsc{QSANNdRA} on the state-of-the-art model more fairly. Finally, we also computed $\bar{\epsilon}$ and $\sigma_\mathrm{\epsilon}$ for the 100 nearest low-redshift quasar neighbours of both ULAS J1120+0641 and ULAS J1342+0928 according to the same procedure as outlined in \ref{sec:2.4}, and compared the results to \textsc{QSANNdRA}.

Figure~\ref{fig:davies_ratio} displays the ratio of \textsc{QSANNdRA}'s relative prediction errors to our calculated relative prediction errors for the original model \citep{2018ApJ...864..142D} and the extended PCA model described above. On average across all target wavelengths, we achieve a 14.2\% improvement in $\bar{\epsilon}$ (left) and a 11.5\% improvement in $\sigma_\mathrm{\epsilon}$ (right) on the full test set as compared to \cite{2018ApJ...864..142D} (shown in gray). When compared to the extended PCA model, we achieve a 6.1\% improvement in $\bar{\epsilon}$ (left) and a 4.9\% improvement in $\sigma_\mathrm{\epsilon}$ (right) on the full test set (shown in magenta) with the least improvement being at the Ly$\alpha$ peak. More interestingly, we can compare the performances of \textsc{QSANNdRA} to the extended PCA method on the QSOs in the SDSS dataset that are most similar to the $z>7$ QSOs that we will use to constrain the high-redshift neutral gas fraction.  To do this, we selected the 100 quasars most similar to ULAS~J1120+0641 and another 100 most similar to ULAS~J1342+0928 based on the red-side of the spectra. On these subsets, \textsc{QSANNdRA}'s performance further improves with respect to the extended PCA method. In particular, on the 100 quasars most similar to ULAS~J1120+0641 (at $z=7.085$, shown in orange), we achieve a 22.1\% improvement in $\bar{\epsilon}$ (left) and a 26.2\% improvement in $\sigma_\mathrm{\epsilon}$ (right). On the 100 nearest neighbours of ULAS~J1342+0928 (at $z=7.5413$, shown in purple), we achieve a 16.8\% improvement in $\bar{\epsilon}$ (left) and a 17.5\% improvement in $\sigma_\mathrm{\epsilon}$ (right).  This experiment indicates that overall, the \textsc{QSANNdRA} is more predictive than the model presented in \cite{2018ApJ...864..142D} as well as its extension and that this is especially true for the two $z>7$ QSOs under consideration.

\section{Application to high-z quasars}\label{sec:3}

Given that we are now confident that our model generalizes to high-redshift objects and performs better than other models in the literature, in this section, we apply \textsc{QSANNdRA} to two high-redshift quasars, namely ULAS~J1120+0641 at $z=7.0851$ \citep{2011Natur.474..616M,2017ApJ...837..146V}, and ULAS~J1342+0928 at $z=7.5413$ \citep{2018Natur.553..473B} to reconstruct their blue-side spectra and constrain the neutral gas fraction at their corresponding redshifts.

\subsection{Reconstructing high-redshift quasar spectra}


In order to apply our algorithm to the high-redshift QSOs, we need to ensure a good fit of the red-side continuum and emission features. Notably, the spectrum of ULAS~J1120+0641 contains a region of poor S/N between $\sim$1660{\AA} and $\sim$1800{\AA}, and between $\sim$2200{\AA} and $\sim$2450{\AA}, and the spectrum of ULAS~J1342+0928 has missing data between $\sim$1570{\AA} and $\sim$1700{\AA}, and between $\sim$2100{\AA} and $\sim$2230{\AA}.

\begin{figure*}
   \centering
   \includegraphics[width=\linewidth,trim={0cm 0.7cm 0cm 0cm},clip]{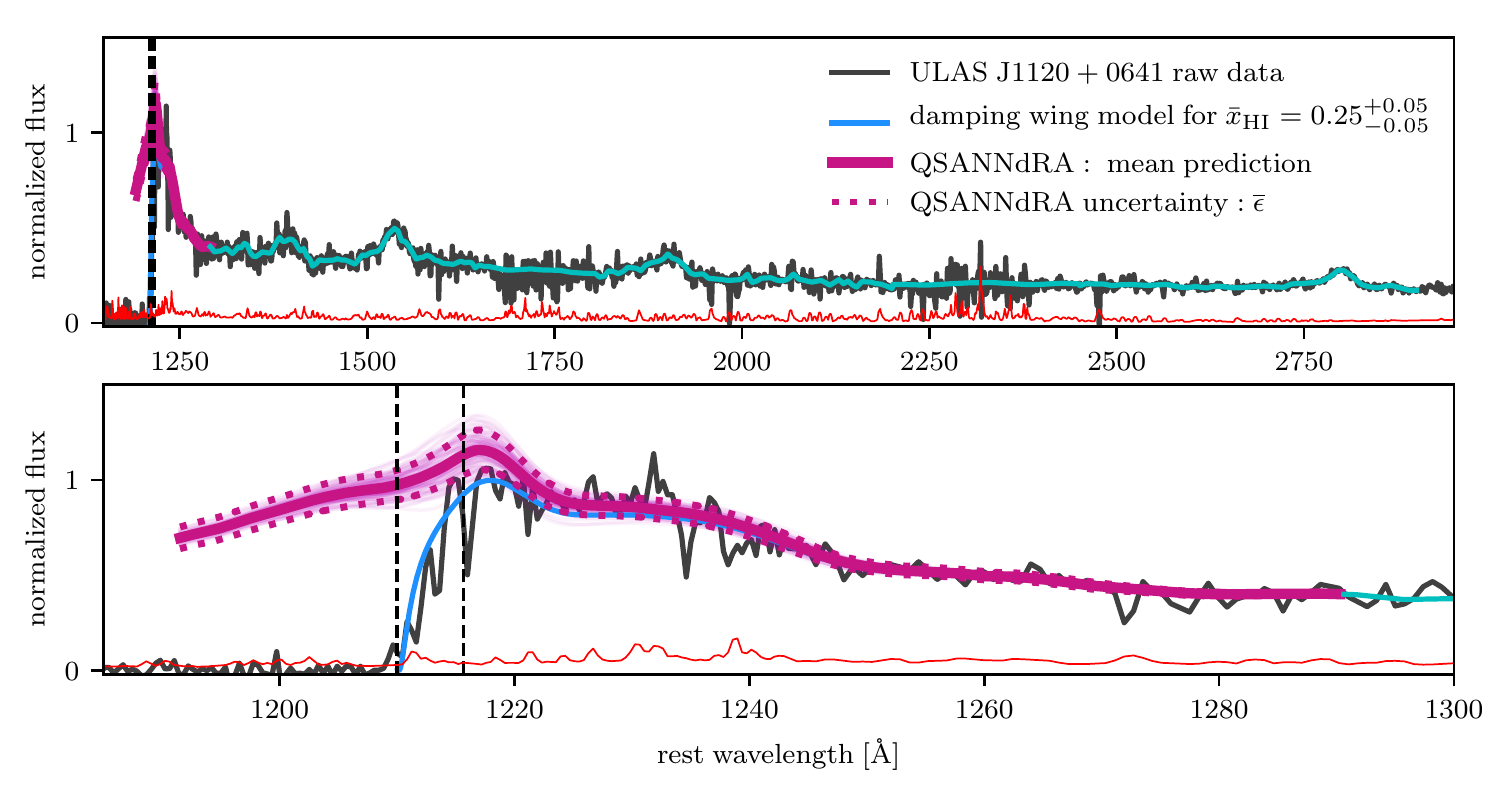}
   \includegraphics[width=\linewidth]{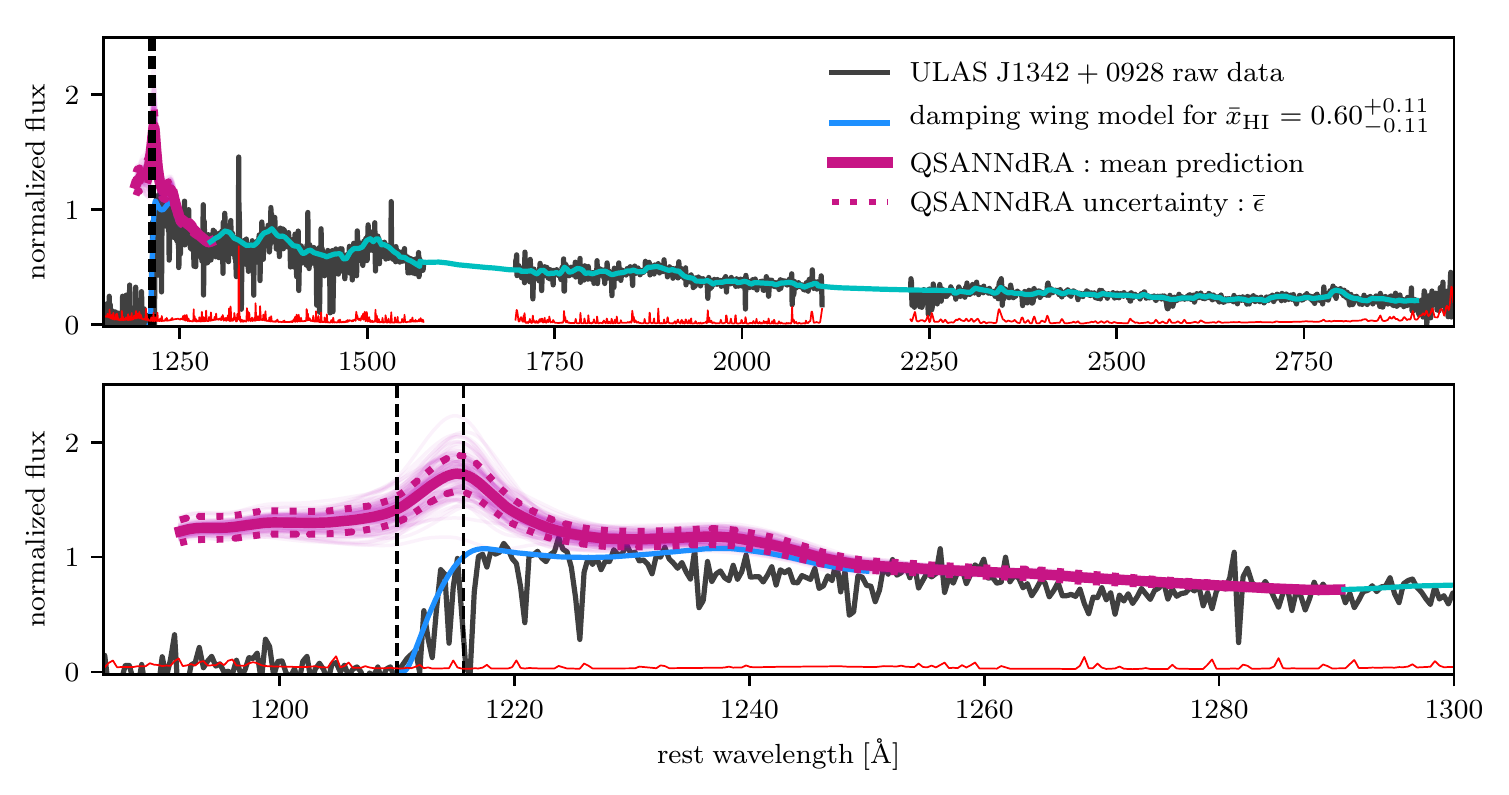}
   \caption{The reconstructed spectrum of ULAS~J1120+0641 (top two panels) and ULAS~J1342+0928 (bottom two panels) with a model of the damping wing. The bottom panel for each quasar shows a close-up view of the Ly$\alpha$ region. The raw data points and their uncertainties are shown in gray and red, respectively. The cyan curve represents our fit of the red-side spectrum.  For  ULAS~J1220+0641, the flux in the poor S/N regions between $\sim$1660{\AA} and $\sim$1800{\AA}, and between $\sim$2200{\AA} and $\sim$2450{\AA} was reconstructed based on the low-redshift QSO spectra.  For ULAS~J1342+0928, the flux in the regions of missing data between $\sim$1570{\AA} and $\sim$1700{\AA}, and between $\sim$2100{\AA} and $\sim$2230{\AA} were also reconstructed based on the low-redshift QSO spectra. The thin light magenta lines show the individual predictions from the 100 NNs with the committee, while the thick magenta line shows the weighted average of these predictions at each wavelength. {We also show the full test set $\bar{\epsilon}$ bounds on our predictions from Figure~\ref{fig:neighbours} as the dotted magenta curves.} The damping wing model is shown in blue and corresponds to $\bar{x}_\mathrm{H\Romannum{1}} = 0.25$ for ULAS~J1220+0641 and $\bar{x}_\mathrm{H\Romannum{1}} = 0.60$ for ULAS~J1342+0928, which was calculated as the weighted average of optimal neutral fractions corresponding to the individual predictions of the 100 networks within the committee. The region between the vertical dashed black lines represents the QSO proximity zone as defined in Section~\ref{sec:2.5}}.
   \label{fig:J1120_pred}
\end{figure*}

To model these parts of the respective spectra as accurately as possible, we took advantage of the correlations between the various features in the spectra again. We trained two very simple NNs for the two cases (with an architecture of 55-20-11 neurons and training in batches of 800 for 400 epochs with the 'elu' activation function), which learnt to predict the poor or missing data based on the remaining parts of the red-side spectra using the training set of low-redshift SDSS QSOs described in Sections~\ref{sec:2.1} and~\ref{sec:2.2}. The resultant predictions had both the mean error and its standard deviation below 2.5\% for all target wavelengths in both cases and displayed a strong generalisability to previously unseen data.

With the fully reconstructed red-side spectrum, we performed the same fitting procedure as outlined in Section~\ref{sec:2.1} and finally applied the trained committee of networks from Section~\ref{sec:2.3} to reconstruct the blue-side spectrum for each $z>7$ QSO.

\begin{figure*}
    \centering
    \includegraphics[]{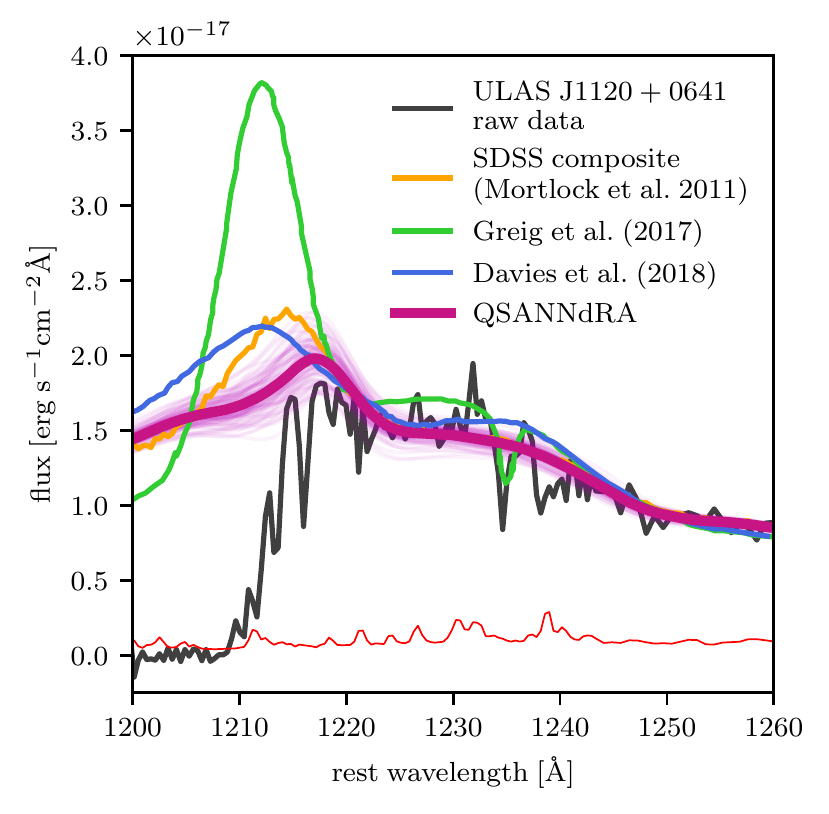}\includegraphics[]{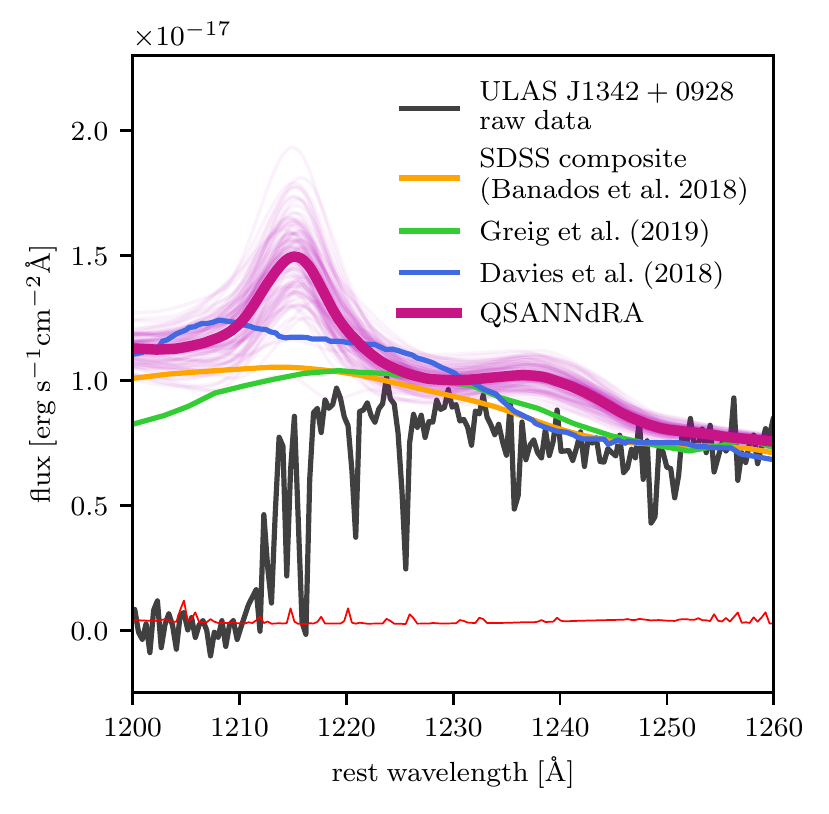}
    \caption{(Left) A comparison of \textsc{QSANNdRA}'s prediction (magenta) for the blue-side spectrum of ULAS~J1120+0641 with existing predictions from the literature, in particular by \protect\cite{2011Natur.474..616M} (orange), \protect\cite{2017MNRAS.466.1814G} (green) and \protect\cite{2018ApJ...864..142D} (blue). We display raw observational data in gray and their corresponding flux uncertainties in red. (Right) A comparison of \textsc{QSANNdRA}'s prediction (magenta) for the blue-side spectrum of ULAS~J1342+0928 with existing predictions from the literature, in particular by \protect\cite{2018Natur.553..473B} (orange), \protect\cite{2019MNRAS.484.5094G} {(green)} and \protect\cite{2018ApJ...864..142D} (blue). We display raw observational data in gray and their corresponding flux uncertainties in red.}
    \label{fig:J1120_comp}
\end{figure*}

In Figure~\ref{fig:J1120_pred}, we show our reconstruction of the continuum spectra of the two $z>7$ QSOs, ULAS~J1220+0641 and ULAS~J1342+0928.  This figure shows the observed quasar spectra in gray along with the uncertainties in each observed data point in red. The cyan curve shows our fit of the red-side spectra. We then show the individual predictions from the 100 NNs in the committee as light magenta curves and emphasise the resultant weighted prediction of the committee as the thick magenta curve. We also include the full test set $\bar{\epsilon}$ uncertainty bounds on our final predictions (see Figure~\ref{fig:neighbours}) as the dotted magenta curves. As expected, the cyan curve that represents the red-side fits the observed data extremely well, even in the regimes where there is low S/N or missing data.  For the $z=7.1$ QSO, ULAS~J1120+0641, the reconstructed continuum has a weak peak near Ly$\alpha$ compared to the observed spectra indicating that limited absorption is occurring at the Ly$\alpha$ peak. In contrast, for the $z=7.5$ QSO, ULAS~J1342+0928, the reconstructed continuum predicts a significantly stronger Ly$\alpha$ peak than what is observed.  Similarly, this QSO also sees more enhanced emission at the location of the NV doublet emission line at a rest-frame wavelength of $\sim$1240{\AA} compared to the slightly lower redshift object. Despite the fact that the SiII emission line at $\sim$1262{\AA} is visible in the error plots for our predictive model, in neither high-redshift QSO do we predict this emission line in reconstruction.  

It is also important to note that all NNs in our model make a prediction that the $z=7.5$ QSO should have had a strong NV line. The fact that this discrepancy between our prediction and the observed flux is much larger than the $\sim 5\%$ uncertainty predicted in Figure~\ref{fig:neighbours} weakens the reliability of the continuum prediction in this case. This discrepancy also impacts the interpretation of the neutral fraction constraint presented in the next section, as more neutral gas is necessary to reconstruct the observed spectrum from our prediction.


In Figure~\ref{fig:J1120_comp}, we compare our reconstructed QSO spectra for the two $z>7$ QSOs with other predictions from the literature.  In the left panel of Figure~\ref{fig:J1120_comp}, we show how our prediction for the blue-side spectrum of ULAS~J1120+0641 (magenta) compares to the predictions based on the SDSS composite (orange) \citep{2011Natur.474..616M}, the covariance matrix approach (green) \citep{2017MNRAS.466.1814G}, and the PCA method (blue) \citep{2018ApJ...864..142D}.  Interestingly, each of the different reconstruction methods gives a different prediction for the intrinsic spectrum.  Our method yields a prediction that is more similar to that of the SDSS composite \citep{2011Natur.474..616M} as well as that predicted in \cite{2018ApJ...864..142D} compared to that predicted in \cite{2017MNRAS.466.1814G}.  As noted earlier, we have predictions from 100 individual NNs that go into our ensemble. None of these 100 NNs predict a Ly$\alpha$ peak that is nearly as strong as that from \cite{2017MNRAS.466.1814G}. In contrast, some of the individual NNs do predict Ly$\alpha$ as strong as that seen in the SDSS composite and the PCA method. Our model for this QSO can in general be categorized as having the weakest Ly$\alpha$ and NV emission, the latter being more consistent with the SDSS composite than that of the PCA method. {Moreover, we also observe a redshifted Ly$\alpha$ peak as compared to the other predictions from the literature. This aspect of our predictions is interesting especially in light of the established correlations between emission line shifts, and could be a consequence of its nearest-neighbour quasars in our standardized PCA space or even potentially hint at a new correlation between emission line profiles. Since obtaining a physical basis for machine learning algorithms is challenging, more investigation needs to be done to better understand this aspect.}

The right panel of Figure~\ref{fig:J1120_comp} displays our prediction for the blue-side spectrum of ULAS J1342+0928 (magenta) in comparison to the predictions based on the SDSS composite (orange) \citep{2018Natur.553..473B}, the covariance matrix approach (green) \citep{2019MNRAS.484.5094G}, and the PCA method (blue) \citep{2018ApJ...864..142D}.  In contrast to the $z=7.1$ QSO where our model predicted the weakest emission lines, for this QSO, \textsc{QSANNdRA} predicts both stronger Ly$\alpha$ and stronger NV emission compared to either the SDSS composite or the PCA method from \cite{2018ApJ...864..142D}.  Hence, our model is in no way biased to predicting either stronger or weaker emission compared to other models in the literature. Some of the 100 individual NNs predict Ly$\alpha$ emission as weak as that reconstructed using the PCA method, however, all NNs in our model do agree on a significantly stronger NV emission and hence a different spectral shape than those predicted by the SDSS composite or the PCA method. Most interesting is how the predictions for the neutral fraction compare given the systematic differences between our model and those from the literature.

\subsection{Constraining the neutral fraction during the Epoch of Reionization}\label{sec:2.5}

In order to determine the neutral fraction based on our reconstructed spectra, we model the damping wing redward of the Gunn-Peterson trough \citep{1965ApJ...142.1633G} according to the analytical model presented in \citet{1998ApJ...501...15M} combined with the Gunn-Peterson optical depth as defined by \citet{2006AJ....132..117F}. {We note that this approach is less sophisticated compared to the methods employed by \cite{2018ApJ...864..143D} and \cite{2017MNRAS.466.4239G,2019MNRAS.484.5094G}, as we do not use simulation-based models of the local high-density environments and gas inflows or outflows.} We use the following cosmological parameters: $h = 0.6766$, $\Omega_\mathrm{m} = 0.3111$ and $\Omega_\mathrm{b} h^2 = 0.02242$ \citep{2018arXiv180706209P}.

The intergalactic medium is modelled as homogeneous and neutral for $z_\mathrm{N} < z < z_\mathrm{S}$, where $z_\mathrm{N}$ is the redshift at which reionization is assumed to be complete, and $z_\mathrm{S}$ is the redshift corresponding to the end of the near zone of the QSO. For $z < z_\mathrm{N}$ the IGM is assumed to be completely ionized. We set  $z_\mathrm{N} = 6$ as compared to a $z_\mathrm{N} = 7$ used by \cite{2018Natur.553..473B}, however, further analysis showed that the model is largely insensitive to the exact value of this redshift (see Appendix~\ref{z_n_app}). Furthermore, the model assumes a fully ionized proximity zone.

We use a common definition for the QSO proximity zone described in the literature \citep{2006AJ....132..117F,2010ApJ...714..834C,2015ApJ...801L..11V,2015MNRAS.454..681K,2018ApJ...864..142D}. We normalised the two fitted high-redshift spectra with respect to the observed flux at Ly$\alpha$, and defined the end of the proximity zone to correspond to the wavelength at which the fitted flux falls below 10\% of the peak value. As an aside, we tested how varying this threshold value impacts the damping wing fit and found that our resultant constraints are insensitive to the exact percentage chosen for the proximity zone definition provided it is $<$ 15\%. With the blue-side predictions at hand for each $z~>~7$ QSO, we then performed a least-squares optimisation to constrain the neutral fraction $\bar{x}_\mathrm{H\Romannum{1}}$ in the damping wing model \citep{1998ApJ...501...15M} by fitting the damped reconstructed continuum  to the smoothed continuum estimate in the wavelength range 1210{\AA} - 1250{\AA}. We used our smoothing algorithm (Appendix~\ref{sec:smooth_app}) to avoid fitting to the absorption features in this wavelength range for both QSOs; however, we note that this works better for the $z~=~7.5$ quasar than for the lower-resolution $z~=~7.1$ QSO. The optimisation procedure was performed on each prediction within the committee individually to obtain a distribution of possible $\bar{x}_\mathrm{H\Romannum{1}}$ for each QSO. Finally, the resulting values of $\bar{x}_\mathrm{H\Romannum{1}}$ were averaged according to the weights of the individual NNs within the committee to obtain a constraint on the neutral fraction at the redshift of each of the two $z>7$ QSOs under consideration. These are shown as the blue lines in Figure~\ref{fig:J1120_pred}. In both high-redshift QSOs, our reconstructed spectrum combined with the damping wing model provides a good representation of the observed data.

We specify the 68\% confidence bounds on our neutral fraction constraints as the standard deviation of the predictions from the 100 NNs in our ensemble. {We find: $\bar{x}_\mathrm{H\Romannum{1}} = 0.25^{+0.05}_{-0.05}$ for ULAS~J1120+0641 at $z = 7.0851$, and $\bar{x}_\mathrm{H\Romannum{1}} = 0.60^{+0.11}_{-0.11}$ for ULAS~J1342+0928 at $z = 7.5413$}.

\begin{figure}
    \centering
    \includegraphics[width=\columnwidth]{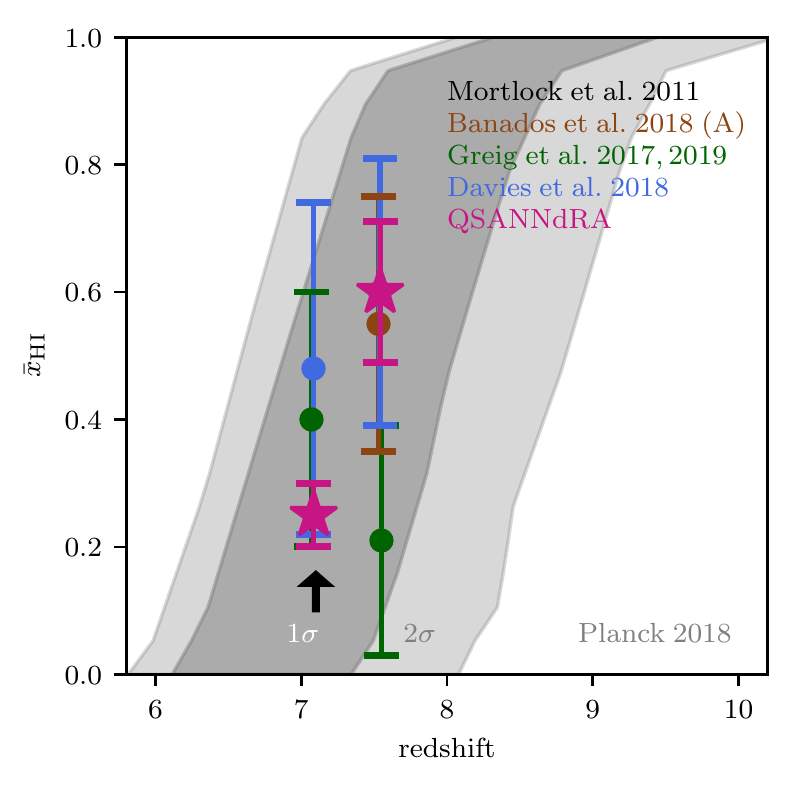}
    \caption{Plot showing the neutral fraction
    constraints published to date. The dark gray and 
    light gray regions show the 1$\sigma$ and 
    2$\sigma$ constraints, respectively, published by 
    the \protect\cite{2018arXiv180706209P} based on CMB 
    observations. We further show constraints based 
    on the damping wing of ULAS J1120+0641 and ULAS 
    J1342+0928 as follows: \protect\cite{2011Natur.474..616M} 
    in black, \protect\cite{2017MNRAS.466.4239G,2019MNRAS.484.5094G} in green, 
    \protect\cite{2018Natur.553..473B} in brown, 
    \protect\cite{2018ApJ...864..143D} in blue, and finally
    our constraints in magenta. We emphasize that due to differences in damping wing models, great care should be taken when directly comparing our constraints to those of \protect\cite{2018ApJ...864..143D} and \protect\cite{2017MNRAS.466.4239G,2019MNRAS.484.5094G}.}
    \label{fig:planck_comp}
\end{figure}

In Figure~\ref{fig:planck_comp}, we compare our neutral fraction predictions with the constraints from the \cite{2018arXiv180706209P} based on CMB observations as well as the estimates from the other models in the literature that modelled the damping-wing of the two high-redshift QSOs.  Our predictions for $\bar{x}_\mathrm{H\Romannum{1}}$ and the $1\sigma$ uncertainties at $z = 7.0851$ and $z = 7.5413$ are well within the $1\sigma$ contours of the estimated prediction from Planck suggesting good agreement between our method and CMB data. Compared to other models in the literature that modelled the damping-wing, our estimates for $\bar{x}_\mathrm{H\Romannum{1}}$ are comparable at $z = 7.5413$, except for \cite{2019MNRAS.484.5094G}, and tend to fall lower at $z = 7.0851$. This is because for the QSO at $z = 7.0851$, we predict a weaker Ly$\alpha$ emission line compared to these other models which means that less neutral hydrogen is needed in the IGM to account for the observed damping wing.

We emphasise that the constraints on the neutral fraction from the literature that are based on the damping-wing use both a different model for reconstructing QSO spectra as well as a different model for how the spectra are processed by the IGM. {We use a simple model of an ionized proximity zone, a completely homogeneous IGM, and a reionization redshift.} This is nearly identical to the Model~A presented in \cite{2018Natur.553..473B}.  Hence, there is a systematic difference in obtaining neutral fraction estimates even after the QSO spectra are reconstructed (see the difference between the three models presented in \citealt{2018Natur.553..473B}). We therefore emphasize that our neutral fraction error bars represent the errors in modelling the intrinsic spectrum and don't reflect any systematic uncertainties due to the differences between homogeneous and inhomogeneous reionization. Nevertheless, with these differences in mind, our model agrees with the others to within $1\sigma$.  In all models, the Universe is neither 100\% neutral at $z = 7.5413$ nor is it 100\% reionized by $z = 7.0851$.  Because of the additional uncertainties in modelling the damping-wing using a more sophisticated model, a significantly larger number of $z>7$ QSOs along multiple lines of sight will be needed to have a statistical estimate for the high-redshift neutral fraction. Nevertheless, our modelling favours a rapid end to reionization.

\section{Summary and conclusions}\label{sec:5}

We have implemented an ensemble of 100 weighted 4-layer fully connected feed-forward neural networks termed \textsc{QSANNdRA} for the purpose of reconstructing the intrinsic high-redshift QSO spectra in the damped region around Ly$\alpha$.  We subsequently use these reconstructions to constrain the neutral gas fraction of the IGM at $z>7$. We trained each individual network in the committee to extract the correlations between the red-side ($1290$\AA~$< \lambda_\mathrm{rest} < 2900$\AA) and the blue-side ($1192$\AA~$< \lambda_\mathrm{rest} < 1290$\AA) spectral features in a sample of 13,703 quasar spectra at redshifts of $2.09 < z < 2.51$ from the SDSS database \citep{2018ApJS..235...42A,2017AJ....154...28B,2013AJ....145...10D,2011AJ....142...72E,2000AJ....120.1579Y}. We applied our trained model to the two highest-redshift QSOs known to date, in particular to ULAS~J1120+0641 at $z=7.0851$ \citep{2011Natur.474..616M,2017ApJ...837..146V}, and ULAS ~J1342+0928 at $z=7.5413$ \citep{2018Natur.553..473B} to reconstruct their continua around Ly$\alpha$. Comparison of our model to the state-of-the-art model reported by \cite{2018ApJ...864..142D} revealed a 14.2\% improvement in the mean relative prediction error across previously unseen low-redshift SDSS QSOs. By extending the PCA model to achieve a fairer comparison, we achieved a 6.1\% improvement in the mean relative prediction errror with the improvement being even more significant for QSOs similar to ULAS~J1120+0641 (22.1\% improvement) and to ULAS~J1342+0928 (16.8\% improvement). Finally, we used our predicted continua and a homogeneous reionization model to constrain the volume-averaged neutral fraction at the redshifts $z=7.0851$ and $z=7.5413$ to be $\bar{x}_\mathrm{H\Romannum{1}} = 0.25^{+0.05}_{-0.05}$ and $\bar{x}_\mathrm{H\Romannum{1}} = 0.60^{+0.11}_{-0.11}$ (with 68\% bounds), respectively.

We emphasize that our constraints use a homogeneous model for the damping wing analysis \citep{1998ApJ...501...15M}, and so our recovered uncertainties on the neutral gas fraction are likely to be under-estimates due to a lack of stochasticity coming from inhomogeneous reionization. A much larger sample of observed high-redshift QSOs will be needed to truly understand this effect. Nonetheless, these constraints are consistent with the literature both for ULAS~J1220+0641 \citep{2011Natur.474..616M,2017MNRAS.466.4239G,2018ApJ...864..142D} and ULAS~J1342+0928 \citep{2018Natur.553..473B,2018ApJ...864..142D}, as well as the estimates from the CMB \citep{2018arXiv180706209P}.  However, our predictions lie on the lower-end of the existing bounds on the neutral fraction at $z=7.0851$ compared to other work that modelled the damping-wing.  This is because for ULAS~J1220+0641, our model predicts weaker intrinsic Ly$\alpha$ emission compared to other models. In addition, the fact that our model predicts a strong NV emission line for ULAS~J1342+0928, which over-predicts the observed flux, affects the interpretation of our neutral fraction constraint.

This is a particularly interesting result, especially in light of the robustness of our prediction model. As \citet{2018ApJ...864..143D} and others \citep{2018Natur.553..473B,2011Natur.474..616M} pointed out, both of these QSOs exhibit outlying spectral features as compared to the low-redshift QSO spectra, the most notable feature being the extremely blueshifted C\Romannum{4} line. Our model is able to capture these outlying features well and thus take their full extent into account when making a prediction. Furthermore, based on our trained autoencoder, the red-side spectral features of the $z=7.1$ QSO are not extreme outliers compared to our training data, and our model is actually expected to perform better on QSOs similar to the ones at $z>7$ compared to the average QSO in SDSS.

The accuracy of our predictions is particularly interesting for another reason. Even though the normalisation and standardization performed in Section~\ref{sec:2.2} was done in order to make the various quasar spectra comparable, it also removed the sensitivity to the Baldwin effect \citep{1977ApJ...214..679B}. Furthermore, quasar luminosity also correlates with emission line shifts \citep{2003ApJ...586...52S,2011AJ....141..167R} possibly because of physical reasons such as orientation \citep{2019MNRAS.487.3305M}. Despite also removing these correlations from our input dataset, our model was still able to achieve low prediction errors, which suggests that these correlations might be implictly contained in other spectral features.

There are two main strengths of our method in the context of high-redshift QSO continua reconstruction. Firstly, we ensure generalizability of \textsc{QSANNdRA}'s performance by constructing the model only on 80\% of QSOs (train subset) from our SDSS training set, while we assess its performance on the remaining, previously unseen, 20\% of QSOs (test subset). The fact that the difference between the performance of our model on the train and test subsets is small suggests that our model can be applied to other, previously unseen QSO without losing accuracy. Secondly, using artificial neural networks is particularly well suited for extracting empirical correlations from large data sets, since it also allows for capturing nonlinear relationships among the various spectral features.

Some concerns might arise about the underlying idea of applying the low-redshift spectral correlations to the high-redshift quasars with unusual spectra, however, these remain to be the best resource available to date for the study of intrinsic spectra of these high-redshift objects.

It is also debatable whether modelling the intervening IGM as homogeneously neutral between the end of the near zones of the QSOs and the reionization redshift is reasonable. Reionization is expected to be patchy (e.g.  \citealt{2006MNRAS.369.1625I,2014ApJ...793..113P,2015PASA...32...45B,2018MNRAS.479.1055B,2019MNRAS.485L..24K}) and therefore we should model a more sophisticated reionization topology than a homogeneous IGM to measure the amount of damping. The highly homogeneous model used in this work \citep{1998ApJ...501...15M} is clearly in contradiction with this theory. However, due to a lack of a statistical sample of QSOs at redshifts relevant to the Epoch of Reionization, it is difficult to establish the details such a model would require to be accurate. Even though the current use of a simplistic model might be justified this way, it should be emphasized that inhomogeneous models are indeed being employed in other work \citep{2011MNRAS.416L..70B,2017MNRAS.466.4239G, 2018ApJ...864..143D,2019MNRAS.484.5094G}.  As more high-redshift QSOs are discovered, our very accurate and generalizable \textsc{QSANNdRA} can be used in the context of a more sophisticated damping-wing model to obtain even better constraints on the high-redshift neutral fraction.

\section*{Acknowledgements}

We would like to thank Dr. Daniel Mortlock for providing the observational data for ULAS J1220+0641, and Dr. Eduardo Ba\~nados for the observational data for ULAS J1342+0928 and valuable feedback. The research of JD and AS is partly funded by Adrian Beecroft and STFC.

Funding for the Sloan Digital Sky Survey IV has been provided by the Alfred P. Sloan Foundation, the U.S. Department of Energy Office of Science, and the Participating Institutions. SDSS-IV acknowledges
support and resources from the Center for High-Performance Computing at
the University of Utah. The SDSS web site is www.sdss.org.

SDSS-IV is managed by the Astrophysical Research Consortium for the 
Participating Institutions of the SDSS Collaboration including the 
Brazilian Participation Group, the Carnegie Institution for Science, 
Carnegie Mellon University, the Chilean Participation Group, the French Participation Group, Harvard-Smithsonian Center for Astrophysics, 
Instituto de Astrof\'isica de Canarias, The Johns Hopkins University, Kavli Institute for the Physics and Mathematics of the Universe (IPMU) / 
University of Tokyo, the Korean Participation Group, Lawrence Berkeley National Laboratory, 
Leibniz Institut f\"ur Astrophysik Potsdam (AIP),  
Max-Planck-Institut f\"ur Astronomie (MPIA Heidelberg), 
Max-Planck-Institut f\"ur Astrophysik (MPA Garching), 
Max-Planck-Institut f\"ur Extraterrestrische Physik (MPE), 
National Astronomical Observatories of China, New Mexico State University, 
New York University, University of Notre Dame, 
Observat\'ario Nacional / MCTI, The Ohio State University, 
Pennsylvania State University, Shanghai Astronomical Observatory, 
United Kingdom Participation Group,
Universidad Nacional Aut\'onoma de M\'exico, University of Arizona, 
University of Colorado Boulder, University of Oxford, University of Portsmouth, 
University of Utah, University of Virginia, University of Washington, University of Wisconsin, 
Vanderbilt University, and Yale University.




\bibliographystyle{mnras}
\bibliography{references} 

\begin{thebibliography}{}
\makeatletter
\relax
\def\mn@urlcharsother{\let\do\@makeother \do\$\do\&\do\#\do\^\do\_\do\%\do\~}
\def\mn@doi{\begingroup\mn@urlcharsother \@ifnextchar [ {\mn@doi@}
  {\mn@doi@[]}}
\def\mn@doi@[#1]#2{\def\@tempa{#1}\ifx\@tempa\@empty \href
  {http://dx.doi.org/#2} {doi:#2}\else \href {http://dx.doi.org/#2} {#1}\fi
  \endgroup}
\def\mn@eprint#1#2{\mn@eprint@#1:#2::\@nil}
\def\mn@eprint@arXiv#1{\href {http://arxiv.org/abs/#1} {{\tt arXiv:#1}}}
\def\mn@eprint@dblp#1{\href {http://dblp.uni-trier.de/rec/bibtex/#1.xml}
  {dblp:#1}}
\def\mn@eprint@#1:#2:#3:#4\@nil{\def\@tempa {#1}\def\@tempb {#2}\def\@tempc
  {#3}\ifx \@tempc \@empty \let \@tempc \@tempb \let \@tempb \@tempa \fi \ifx
  \@tempb \@empty \def\@tempb {arXiv}\fi \@ifundefined
  {mn@eprint@\@tempb}{\@tempb:\@tempc}{\expandafter \expandafter \csname
  mn@eprint@\@tempb\endcsname \expandafter{\@tempc}}}

\bibitem[\protect\citeauthoryear{{Abolfathi} et~al.,}{{Abolfathi}
  et~al.}{2018}]{2018ApJS..235...42A}
{Abolfathi} B.,  et~al., 2018, \mn@doi [The Astrophysical Journal Supplement
  Series] {10.3847/1538-4365/aa9e8a}, \href
  {http://adsabs.harvard.edu/abs/2018ApJS..235...42A} {235, 42}

\bibitem[\protect\citeauthoryear{{Ba{\~n}ados} et~al.,}{{Ba{\~n}ados}
  et~al.}{2018}]{2018Natur.553..473B}
{Ba{\~n}ados} E.,  et~al., 2018, \mn@doi [\nat] {10.1038/nature25180}, \href
  {https://ui.adsabs.harvard.edu/abs/2018Natur.553..473B} {553, 473}

\bibitem[\protect\citeauthoryear{{Baldwin}}{{Baldwin}}{1977}]{1977ApJ...214..679B}
{Baldwin} J.~A.,  1977, \mn@doi [\apj] {10.1086/155294}, \href
  {https://ui.adsabs.harvard.edu/abs/1977ApJ...214..679B} {214, 679}

\bibitem[\protect\citeauthoryear{{Becker}, {Bolton}  \& {Lidz}}{{Becker}
  et~al.}{2015}]{2015PASA...32...45B}
{Becker} G.~D.,  {Bolton} J.~S.,   {Lidz} A.,  2015, \mn@doi [\pasa]
  {10.1017/pasa.2015.45}, \href
  {https://ui.adsabs.harvard.edu/abs/2015PASA...32...45B} {32, e045}

\bibitem[\protect\citeauthoryear{{Blanton} et~al.,}{{Blanton}
  et~al.}{2017}]{2017AJ....154...28B}
{Blanton} M.~R.,  et~al., 2017, \mn@doi [\aj] {10.3847/1538-3881/aa7567}, \href
  {http://adsabs.harvard.edu/abs/2017AJ....154...28B} {154, 28}

\bibitem[\protect\citeauthoryear{{Bolton}, {Haehnelt}, {Warren}, {Hewett},
  {Mortlock}, {Venemans}, {McMahon}  \& {Simpson}}{{Bolton}
  et~al.}{2011}]{2011MNRAS.416L..70B}
{Bolton} J.~S.,  {Haehnelt} M.~G.,  {Warren} S.~J.,  {Hewett} P.~C.,
  {Mortlock} D.~J.,  {Venemans} B.~P.,  {McMahon} R.~G.,   {Simpson} C.,  2011,
  \mn@doi [\mnras] {10.1111/j.1745-3933.2011.01100.x}, \href
  {https://ui.adsabs.harvard.edu/abs/2011MNRAS.416L..70B} {416, L70}

\bibitem[\protect\citeauthoryear{{Boroson} \& {Green}}{{Boroson} \&
  {Green}}{1992}]{1992ApJS...80..109B}
{Boroson} T.~A.,  {Green} R.~F.,  1992, \mn@doi [\apjs] {10.1086/191661}, \href
  {https://ui.adsabs.harvard.edu/abs/1992ApJS...80..109B} {80, 109}

\bibitem[\protect\citeauthoryear{{Bosman}, {Fan}, {Jiang}, {Reed}, {Matsuoka},
  {Becker}  \& {Haehnelt}}{{Bosman} et~al.}{2018}]{2018MNRAS.479.1055B}
{Bosman} S. E.~I.,  {Fan} X.,  {Jiang} L.,  {Reed} S.,  {Matsuoka} Y.,
  {Becker} G.,   {Haehnelt} M.,  2018, \mn@doi [\mnras]
  {10.1093/mnras/sty1344}, \href
  {https://ui.adsabs.harvard.edu/abs/2018MNRAS.479.1055B} {479, 1055}

\bibitem[\protect\citeauthoryear{Breiman}{Breiman}{2001}]{Breiman2001}
Breiman L.,  2001, \mn@doi [Machine Learning] {10.1023/A:1010933404324}, 45, 5

\bibitem[\protect\citeauthoryear{{Carilli} et~al.,}{{Carilli}
  et~al.}{2010}]{2010ApJ...714..834C}
{Carilli} C.~L.,  et~al., 2010, \mn@doi [\apj] {10.1088/0004-637X/714/1/834},
  \href {https://ui.adsabs.harvard.edu/abs/2010ApJ...714..834C} {714, 834}

\bibitem[\protect\citeauthoryear{Chollet et~al.}{Chollet
  et~al.}{2015}]{chollet2015keras}
Chollet F.,  et~al., 2015, Keras, \url{https://keras.io}

\bibitem[\protect\citeauthoryear{{Davies} et~al.,}{{Davies}
  et~al.}{2018a}]{2018ApJ...864..142D}
{Davies} F.~B.,  et~al., 2018a, \mn@doi [\apj] {10.3847/1538-4357/aad6dc},
  \href {https://ui.adsabs.harvard.edu/abs/2018ApJ...864..142D} {864, 142}

\bibitem[\protect\citeauthoryear{{Davies} et~al.,}{{Davies}
  et~al.}{2018b}]{2018ApJ...864..143D}
{Davies} F.~B.,  et~al., 2018b, \mn@doi [\apj] {10.3847/1538-4357/aad7f8},
  \href {https://ui.adsabs.harvard.edu/abs/2018ApJ...864..143D} {864, 143}

\bibitem[\protect\citeauthoryear{{Dawson} et~al.,}{{Dawson}
  et~al.}{2013}]{2013AJ....145...10D}
{Dawson} K.~S.,  et~al., 2013, \mn@doi [\aj] {10.1088/0004-6256/145/1/10},
  \href {http://adsabs.harvard.edu/abs/2013AJ....145...10D} {145, 10}

\bibitem[\protect\citeauthoryear{{Dawson} et~al.,}{{Dawson}
  et~al.}{2016}]{2016AJ....151...44D}
{Dawson} K.~S.,  et~al., 2016, \mn@doi [\aj] {10.3847/0004-6256/151/2/44},
  \href {http://adsabs.harvard.edu/abs/2016AJ....151...44D} {151, 44}

\bibitem[\protect\citeauthoryear{Dietterich}{Dietterich}{2000}]{dietterich2000ensemble}
Dietterich T.~G.,  2000, in International workshop on multiple classifier
  systems. pp 1--15

\bibitem[\protect\citeauthoryear{{Eilers}, {Davies}, {Hennawi}, {Prochaska},
  {Luki{\'c}}  \& {Mazzucchelli}}{{Eilers} et~al.}{2017}]{2017ApJ...840...24E}
{Eilers} A.-C.,  {Davies} F.~B.,  {Hennawi} J.~F.,  {Prochaska} J.~X.,
  {Luki{\'c}} Z.,   {Mazzucchelli} C.,  2017, \mn@doi [\apj]
  {10.3847/1538-4357/aa6c60}, \href
  {https://ui.adsabs.harvard.edu/abs/2017ApJ...840...24E} {840, 24}

\bibitem[\protect\citeauthoryear{{Eilers}, {Davies}  \& {Hennawi}}{{Eilers}
  et~al.}{2018}]{2018ApJ...864...53E}
{Eilers} A.-C.,  {Davies} F.~B.,   {Hennawi} J.~F.,  2018, \mn@doi [\apj]
  {10.3847/1538-4357/aad4fd}, \href
  {https://ui.adsabs.harvard.edu/abs/2018ApJ...864...53E} {864, 53}

\bibitem[\protect\citeauthoryear{{Eilers}, {Hennawi}, {Davies}  \&
  {O{\~n}orbe}}{{Eilers} et~al.}{2019}]{2019ApJ...881...23E}
{Eilers} A.-C.,  {Hennawi} J.~F.,  {Davies} F.~B.,   {O{\~n}orbe} J.,  2019,
  \mn@doi [\apj] {10.3847/1538-4357/ab2b3f}, \href
  {https://ui.adsabs.harvard.edu/abs/2019ApJ...881...23E} {881, 23}

\bibitem[\protect\citeauthoryear{{Eisenstein} et~al.,}{{Eisenstein}
  et~al.}{2011}]{2011AJ....142...72E}
{Eisenstein} D.~J.,  et~al., 2011, \mn@doi [\aj] {10.1088/0004-6256/142/3/72},
  \href {http://adsabs.harvard.edu/abs/2011AJ....142...72E} {142, 72}

\bibitem[\protect\citeauthoryear{{Fan} et~al.,}{{Fan}
  et~al.}{2006}]{2006AJ....132..117F}
{Fan} X.,  et~al., 2006, \mn@doi [\aj] {10.1086/504836}, \href
  {https://ui.adsabs.harvard.edu/abs/2006AJ....132..117F} {132, 117}

\bibitem[\protect\citeauthoryear{Fischler \& Bolles}{Fischler \&
  Bolles}{1981}]{fischler1981random}
Fischler M.~A.,  Bolles R.~C.,  1981, Communications of the ACM, 24, 381

\bibitem[\protect\citeauthoryear{{Francis}, {Hewett}, {Foltz}  \&
  {Chaffee}}{{Francis} et~al.}{1992}]{1992ApJ...398..476F}
{Francis} P.~J.,  {Hewett} P.~C.,  {Foltz} C.~B.,   {Chaffee} F.~H.,  1992,
  \mn@doi [\apj] {10.1086/171870}, \href
  {https://ui.adsabs.harvard.edu/abs/1992ApJ...398..476F} {398, 476}

\bibitem[\protect\citeauthoryear{G{\'e}ron}{G{\'e}ron}{2017}]{geron2017hands}
G{\'e}ron A.,  2017, Hands-on machine learning with Scikit-Learn and
  TensorFlow: concepts, tools, and techniques to build intelligent systems.
O'Reilly Media, Inc.

\bibitem[\protect\citeauthoryear{{Greig}, {Mesinger}, {McGreer}, {Gallerani}
  \& {Haiman}}{{Greig} et~al.}{2017a}]{2017MNRAS.466.1814G}
{Greig} B.,  {Mesinger} A.,  {McGreer} I.~D.,  {Gallerani} S.,   {Haiman} Z.,
  2017a, \mn@doi [\mnras] {10.1093/mnras/stw3210}, \href
  {https://ui.adsabs.harvard.edu/abs/2017MNRAS.466.1814G} {466, 1814}

\bibitem[\protect\citeauthoryear{{Greig}, {Mesinger}, {Haiman}  \&
  {Simcoe}}{{Greig} et~al.}{2017b}]{2017MNRAS.466.4239G}
{Greig} B.,  {Mesinger} A.,  {Haiman} Z.,   {Simcoe} R.~A.,  2017b, \mn@doi
  [\mnras] {10.1093/mnras/stw3351}, \href
  {https://ui.adsabs.harvard.edu/abs/2017MNRAS.466.4239G} {466, 4239}

\bibitem[\protect\citeauthoryear{{Greig}, {Mesinger}  \& {Ba{\~n}ados}}{{Greig}
  et~al.}{2019}]{2019MNRAS.484.5094G}
{Greig} B.,  {Mesinger} A.,   {Ba{\~n}ados} E.,  2019, \mn@doi [\mnras]
  {10.1093/mnras/stz230}, \href
  {https://ui.adsabs.harvard.edu/abs/2019MNRAS.484.5094G} {484, 5094}

\bibitem[\protect\citeauthoryear{{Gunn} \& {Peterson}}{{Gunn} \&
  {Peterson}}{1965}]{1965ApJ...142.1633G}
{Gunn} J.~E.,  {Peterson} B.~A.,  1965, \mn@doi [\apj] {10.1086/148444}, \href
  {https://ui.adsabs.harvard.edu/abs/1965ApJ...142.1633G} {142, 1633}

\bibitem[\protect\citeauthoryear{{Hewett} \& {Wild}}{{Hewett} \&
  {Wild}}{2010}]{2010MNRAS.405.2302H}
{Hewett} P.~C.,  {Wild} V.,  2010, \mn@doi [\mnras]
  {10.1111/j.1365-2966.2010.16648.x}, \href
  {https://ui.adsabs.harvard.edu/abs/2010MNRAS.405.2302H} {405, 2302}

\bibitem[\protect\citeauthoryear{{Iliev}, {Mellema}, {Pen}, {Merz}, {Shapiro}
  \& {Alvarez}}{{Iliev} et~al.}{2006}]{2006MNRAS.369.1625I}
{Iliev} I.~T.,  {Mellema} G.,  {Pen} U.~L.,  {Merz} H.,  {Shapiro} P.~R.,
  {Alvarez} M.~A.,  2006, \mn@doi [\mnras] {10.1111/j.1365-2966.2006.10502.x},
  \href {https://ui.adsabs.harvard.edu/abs/2006MNRAS.369.1625I} {369, 1625}

\bibitem[\protect\citeauthoryear{Jones, Oliphant, Peterson  et~al.}{Jones
  et~al.}{2001}]{scipy}
Jones E.,  Oliphant T.,  Peterson P.,   et~al., 2001, {SciPy}: Open source
  scientific tools for {Python}, \url {http://www.scipy.org/}

\bibitem[\protect\citeauthoryear{{Keating}, {Haehnelt}, {Cantalupo}  \&
  {Puchwein}}{{Keating} et~al.}{2015}]{2015MNRAS.454..681K}
{Keating} L.~C.,  {Haehnelt} M.~G.,  {Cantalupo} S.,   {Puchwein} E.,  2015,
  \mn@doi [\mnras] {10.1093/mnras/stv2020}, \href
  {https://ui.adsabs.harvard.edu/abs/2015MNRAS.454..681K} {454, 681}

\bibitem[\protect\citeauthoryear{{Kulkarni}, {Keating}, {Haehnelt}, {Bosman},
  {Puchwein}, {Chardin}  \& {Aubert}}{{Kulkarni}
  et~al.}{2019}]{2019MNRAS.485L..24K}
{Kulkarni} G.,  {Keating} L.~C.,  {Haehnelt} M.~G.,  {Bosman} S. E.~I.,
  {Puchwein} E.,  {Chardin} J.,   {Aubert} D.,  2019, \mn@doi [\mnras]
  {10.1093/mnrasl/slz025}, \href
  {https://ui.adsabs.harvard.edu/abs/2019MNRAS.485L..24K} {485, L24}

\bibitem[\protect\citeauthoryear{{Meyer}, {Bosman}  \& {Ellis}}{{Meyer}
  et~al.}{2019}]{2019MNRAS.487.3305M}
{Meyer} R.~A.,  {Bosman} S. E.~I.,   {Ellis} R.~S.,  2019, \mn@doi [\mnras]
  {10.1093/mnras/stz1504}, \href
  {https://ui.adsabs.harvard.edu/abs/2019MNRAS.487.3305M} {487, 3305}

\bibitem[\protect\citeauthoryear{{Miralda-Escud{\'e}}}{{Miralda-Escud{\'e}}}{1998}]{1998ApJ...501...15M}
{Miralda-Escud{\'e}} J.,  1998, \mn@doi [\apj] {10.1086/305799}, \href
  {https://ui.adsabs.harvard.edu/abs/1998ApJ...501...15M} {501, 15}

\bibitem[\protect\citeauthoryear{{Mortlock}}{{Mortlock}}{2016}]{2016ASSL..423..187M}
{Mortlock} D.,  2016, in {Mesinger} A.,  ed.,  Astrophysics and Space Science
  Library Vol. 423, Understanding the Epoch of Cosmic Reionization: Challenges
  and Progress. p.~187 (\mn@eprint {arXiv} {1511.01107}),
  \mn@doi{10.1007/978-3-319-21957-8_7}

\bibitem[\protect\citeauthoryear{{Mortlock} et~al.,}{{Mortlock}
  et~al.}{2011}]{2011Natur.474..616M}
{Mortlock} D.~J.,  et~al., 2011, \mn@doi [\nat] {10.1038/nature10159}, \href
  {https://ui.adsabs.harvard.edu/abs/2011Natur.474..616M} {474, 616}

\bibitem[\protect\citeauthoryear{{P{\^a}ris} et~al.,}{{P{\^a}ris}
  et~al.}{2011}]{2011A&A...530A..50P}
{P{\^a}ris} I.,  et~al., 2011, \mn@doi [\aap] {10.1051/0004-6361/201016233},
  \href {https://ui.adsabs.harvard.edu/abs/2011A&A...530A..50P} {530, A50}

\bibitem[\protect\citeauthoryear{{P{\^a}ris} et~al.,}{{P{\^a}ris}
  et~al.}{2018}]{2018A&A...613A..51P}
{P{\^a}ris} I.,  et~al., 2018, \mn@doi [\aap] {10.1051/0004-6361/201732445},
  \href {https://ui.adsabs.harvard.edu/abs/2018A&A...613A..51P} {613, A51}

\bibitem[\protect\citeauthoryear{{Pedregosa} et~al.,}{{Pedregosa}
  et~al.}{2012}]{2012arXiv1201.0490P}
{Pedregosa} F.,  et~al., 2012, arXiv e-prints, \href
  {https://ui.adsabs.harvard.edu/abs/2012arXiv1201.0490P} {p. arXiv:1201.0490}

\bibitem[\protect\citeauthoryear{{Pentericci} et~al.,}{{Pentericci}
  et~al.}{2014}]{2014ApJ...793..113P}
{Pentericci} L.,  et~al., 2014, \mn@doi [\apj] {10.1088/0004-637X/793/2/113},
  \href {https://ui.adsabs.harvard.edu/abs/2014ApJ...793..113P} {793, 113}

\bibitem[\protect\citeauthoryear{{Planck Collaboration} et~al.,}{{Planck
  Collaboration} et~al.}{2018}]{2018arXiv180706209P}
{Planck Collaboration} et~al., 2018, arXiv e-prints, \href
  {https://ui.adsabs.harvard.edu/abs/2018arXiv180706209P} {p. arXiv:1807.06209}

\bibitem[\protect\citeauthoryear{{Richards} et~al.,}{{Richards}
  et~al.}{2011}]{2011AJ....141..167R}
{Richards} G.~T.,  et~al., 2011, \mn@doi [\aj] {10.1088/0004-6256/141/5/167},
  \href {https://ui.adsabs.harvard.edu/abs/2011AJ....141..167R} {141, 167}

\bibitem[\protect\citeauthoryear{{Shang}, {Wills}, {Robinson}, {Wills}, {Laor},
  {Xie}  \& {Yuan}}{{Shang} et~al.}{2003}]{2003ApJ...586...52S}
{Shang} Z.,  {Wills} B.~J.,  {Robinson} E.~L.,  {Wills} D.,  {Laor} A.,  {Xie}
  B.,   {Yuan} J.,  2003, \mn@doi [\apj] {10.1086/367638}, \href
  {https://ui.adsabs.harvard.edu/abs/2003ApJ...586...52S} {586, 52}

\bibitem[\protect\citeauthoryear{{Shang}, {Wills}, {Wills}  \&
  {Brotherton}}{{Shang} et~al.}{2007}]{2007AJ....134..294S}
{Shang} Z.,  {Wills} B.~J.,  {Wills} D.,   {Brotherton} M.~S.,  2007, \mn@doi
  [\aj] {10.1086/518505}, \href
  {https://ui.adsabs.harvard.edu/abs/2007AJ....134..294S} {134, 294}

\bibitem[\protect\citeauthoryear{{Shen} et~al.,}{{Shen}
  et~al.}{2016}]{2016ApJ...831....7S}
{Shen} Y.,  et~al., 2016, \mn@doi [\apj] {10.3847/0004-637X/831/1/7}, \href
  {https://ui.adsabs.harvard.edu/abs/2016ApJ...831....7S} {831, 7}

\bibitem[\protect\citeauthoryear{{Stoughton} et~al.,}{{Stoughton}
  et~al.}{2002}]{2002AJ....123..485S}
{Stoughton} C.,  et~al., 2002, \mn@doi [\aj] {10.1086/324741}, \href
  {https://ui.adsabs.harvard.edu/abs/2002AJ....123..485S} {123, 485}

\bibitem[\protect\citeauthoryear{{Suzuki}}{{Suzuki}}{2006}]{2006ApJS..163..110S}
{Suzuki} N.,  2006, \mn@doi [\apjs] {10.1086/499272}, \href
  {https://ui.adsabs.harvard.edu/abs/2006ApJS..163..110S} {163, 110}

\bibitem[\protect\citeauthoryear{{Suzuki}, {Tytler}, {Kirkman}, {O'Meara}  \&
  {Lubin}}{{Suzuki} et~al.}{2005}]{2005ApJ...618..592S}
{Suzuki} N.,  {Tytler} D.,  {Kirkman} D.,  {O'Meara} J.~M.,   {Lubin} D.,
  2005, \mn@doi [\apj] {10.1086/426062}, \href
  {https://ui.adsabs.harvard.edu/abs/2005ApJ...618..592S} {618, 592}

\bibitem[\protect\citeauthoryear{{Venemans} et~al.,}{{Venemans}
  et~al.}{2015}]{2015ApJ...801L..11V}
{Venemans} B.~P.,  et~al., 2015, \mn@doi [\apj] {10.1088/2041-8205/801/1/L11},
  \href {https://ui.adsabs.harvard.edu/abs/2015ApJ...801L..11V} {801, L11}

\bibitem[\protect\citeauthoryear{{Venemans} et~al.,}{{Venemans}
  et~al.}{2017a}]{2017ApJ...837..146V}
{Venemans} B.~P.,  et~al., 2017a, \mn@doi [\apj] {10.3847/1538-4357/aa62ac},
  \href {https://ui.adsabs.harvard.edu/abs/2017ApJ...837..146V} {837, 146}

\bibitem[\protect\citeauthoryear{{Venemans} et~al.,}{{Venemans}
  et~al.}{2017b}]{2017ApJ...851L...8V}
{Venemans} B.~P.,  et~al., 2017b, \mn@doi [\apjl] {10.3847/2041-8213/aa943a},
  \href {https://ui.adsabs.harvard.edu/abs/2017ApJ...851L...8V} {851, L8}

\bibitem[\protect\citeauthoryear{{Yip} et~al.,}{{Yip}
  et~al.}{2004}]{2004AJ....128.2603Y}
{Yip} C.~W.,  et~al., 2004, \mn@doi [\aj] {10.1086/425626}, \href
  {https://ui.adsabs.harvard.edu/abs/2004AJ....128.2603Y} {128, 2603}

\bibitem[\protect\citeauthoryear{{York} et~al.,}{{York}
  et~al.}{2000}]{2000AJ....120.1579Y}
{York} D.~G.,  et~al., 2000, \mn@doi [\aj] {10.1086/301513}, \href
  {http://adsabs.harvard.edu/abs/2000AJ....120.1579Y} {120, 1579}

\makeatother
\end{thebibliography}



\appendix
\vspace{-10pt}
\section{S/N cut analysis}\label{SN_app}

This Appendix reports the impact that different S/N cuts in the cleaning stage can have on the overall performance of our model.

\begin{figure}
    \centering
    \includegraphics[width=\columnwidth]{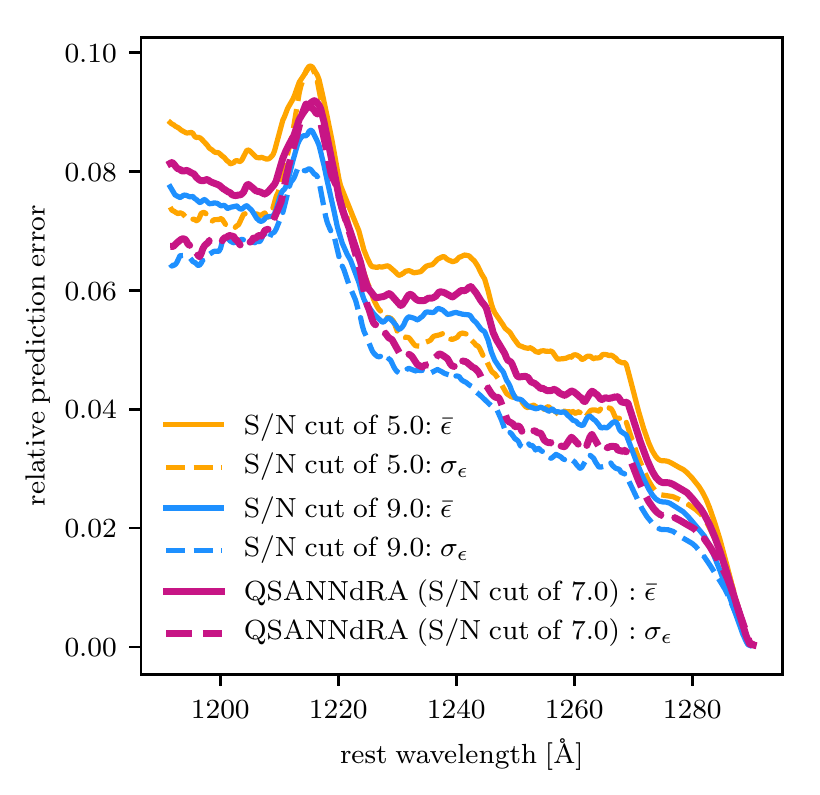}
    \caption{Dependence of \textsc{QSANNdRA}'s performance on the S/N cut threshold in the data pre-processing stage shown in terms of $\bar{\epsilon}$ (solid lines) and $\sigma_\mathrm{\epsilon}$ (dashed lines). The thick magenta curves display \textsc{QSANNdRA}'s performance reported in the main text, while the orange and the blue curves display the model's performance when trained on a dataset of low-redshift quasars with S/N $\geq$ 5 and S/N $\geq$ 9, respectively. }
    \label{fig:SN_cut_analysis}
\end{figure}

While \textsc{QSANNdRA} has been trained on a low-redshift data set composed of QSOs with ${\rm S/N}\geq7$, we aim to understand how changing this parameter might impact its performance and predictions. In order to investigate this, we created two more training data sets with a S/N cut of 5 and 9, respectively, and then retrained our model on these new data sets while keeping everything else the same. Figure~\ref{fig:SN_cut_analysis} shows the test set performance of the two new models as well as the baseline performance reported in the main text of this paper. The performance arising from a S/N cut of 5 is shown in orange, the one arising from a S/N cut of 9 is shown in blue, and the performance of the fiducial model, \textsc{QSANNdRA}, is shown in magenta. For each case, we show $\bar{\epsilon}$ as a solid curve and $\sigma_\mathrm{\epsilon}$ as a dashed curve.

While it seems that the performance improves for higher S/N cuts, the conclusion from this analysis has to be drawn carefully. For instance, while it is easier for the networks to learn the underlying spectral correlations in a less noisy data set, using a higher S/N cut comes at the cost of a smaller training data set. In particular, while the original ${\rm S/N}\geq 7$ data set has 13,703 quasars, the ${\rm S/N}\geq5$ data set has 22,753 quasars and the ${\rm S/N}\geq9$ has only 9,229 quasars. These two aspects ultimately need to be balanced to achieve both a considerably small prediction uncertainty as well as good generalizability.

\begin{figure}
    \centering
    \includegraphics[]{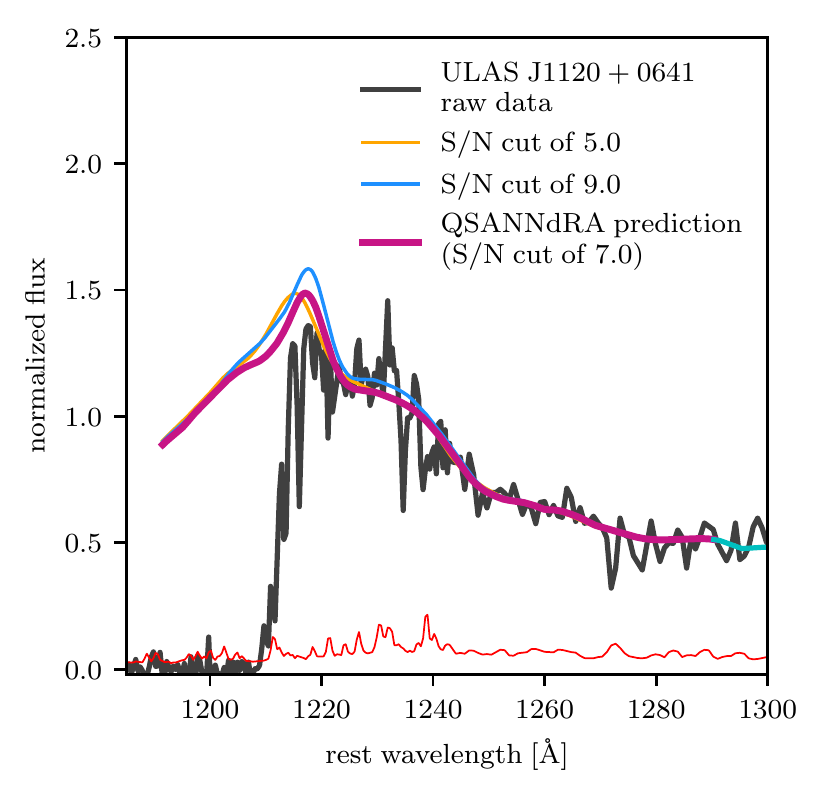}
    \includegraphics[]{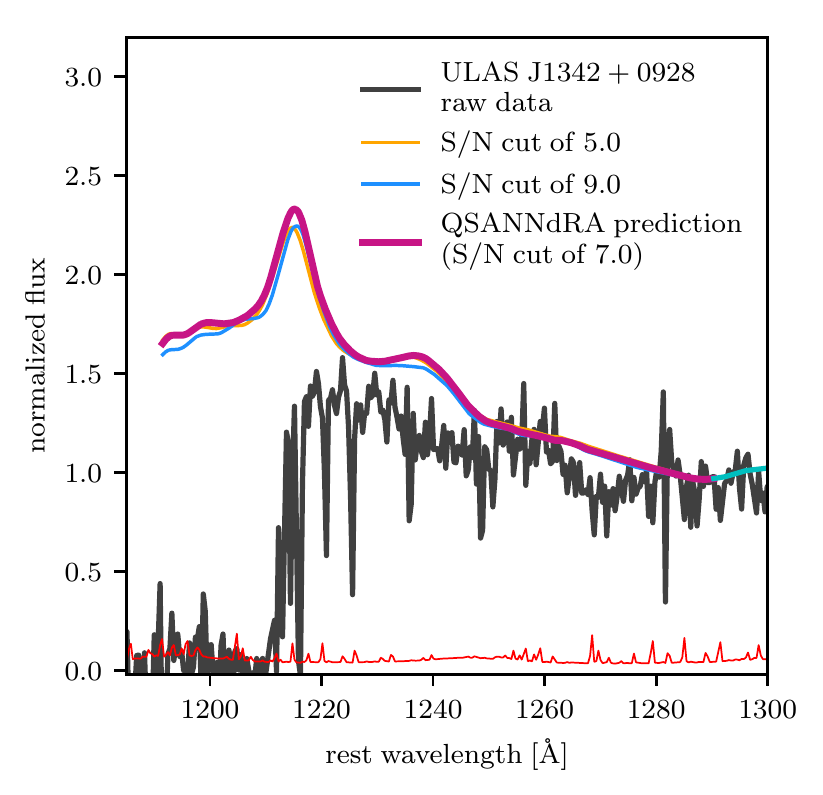}
    \caption{A comparison of the predicted fluxes based on ${\rm S/N}\geq5$ (orange) and ${\rm S/N}\geq9$ (blue) data sets of low-z quasars to the main predictions of \textsc{QSANNdRA} based on a ${\rm S/N}\geq7$ (magenta) data set of low-z quasars for ULAS J1120+0641 (top) and ULAS J1342+0928 (bottom).}
    \label{fig:SN_preds}
\end{figure}

\begin{figure}
    \centering
    \includegraphics[width=\columnwidth]{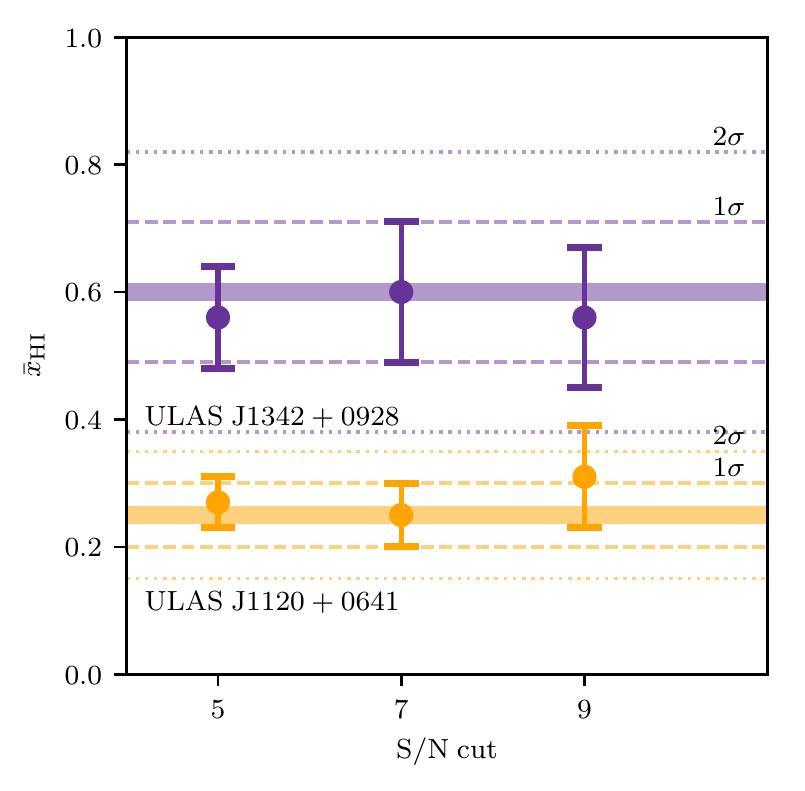}
    \caption{A comparison of the resultant neutral fraction constraints for ULAS J1120+0641 (orange) and ULAS J1342+0928 (purple) arising from a S/N cut of 5, 7 and 9 on the low-redshift SDSS quasars used for the construction of \textsc{QSANNdRA}. The horizontal solid lines depict the baseline neutral fraction constraints  $\bar{x}_\mathrm{H\Romannum{1}} = 0.25$ for ULAS J1120+0641 and $\bar{x}_\mathrm{H\Romannum{1}} = 0.60$ for ULAS J1342+0928, while the dashed and dotted lines represent the posterior 1$\sigma$ and 2$\sigma$ uncertainty bounds, respectively.}
    \label{fig:SN_xHI}
\end{figure}

We further applied the two retrained models to the two high-z quasars, namely ULAS~J1120+0641 at $z=7.0851$ \citep{2011Natur.474..616M}, and ULAS~J1342+0928 at $z=7.5413$ \citep{2018Natur.553..473B}, and used their predictions to constrain the neutral fraction as outlined in the main text. Figure~\ref{fig:SN_preds} displays the resultant predictions for ULAS J1120+0641 (top panel) and ULAS J1342+0928 (bottom panel) based on the ${\rm S/N}\geq5$ and ${\rm S/N}\geq9$ data sets in orange and blue, respectively, and also shows our main prediction in magenta for comparison. We observe a slight increase in the predicted flux for the $z=7.1$ QSO and a slight decrease in the predicted flux for the $z=7.5$ QSO, which then translates into slightly higher neutral fraction constraints for ULAS J1120+0641 and slightly lower neutral fraction contraints for ULAS J1342+0928 (Figure~\ref{fig:SN_xHI}).

\vspace{-10pt}
\section{Smoothing algorithm}\label{sec:smooth_app}

\begin{figure*}
    \centering
    \includegraphics[width=\linewidth]{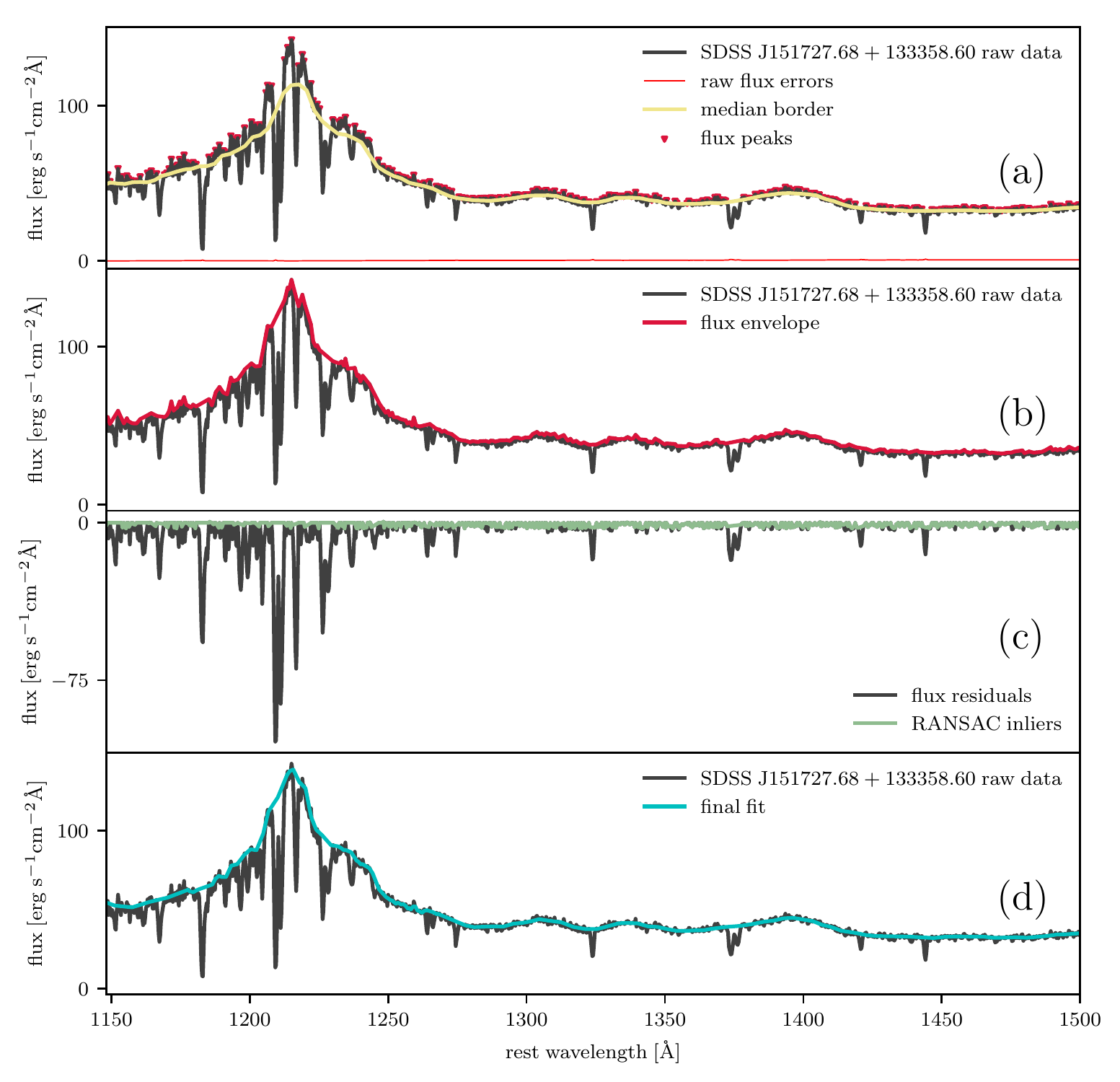}
    \caption{An illustration of our smoothing algorithm. We first compute a running median with a bin size of 50 data points to capture the main continuum and emission features in the spectrum (a, yellow). We then perform a peak-finding procedure using the SciPy Python library \citep{scipy} above the aforementioned running median border (a, red points) and interpolate the peaks to construct and upper envelope of the spectrum (b, red). This envelope is then subtracted from the spectrum, thus linearizing our data (c, black). We then applied the RANSAC regressor algorithm \citep{fischler1981random} from the Scikit-Learn Python package \citep{2012arXiv1201.0490P} on the residuals, thus rejecting most absorption features in the spectrum. The data points which were flagged as inliers by RANSAC (c, green) were interpolated and smoothed by computing a running median with a bin size of 20, thus creating the final flux fit of each spectrum (d, cyan).}
    \label{fig:smoothing}
\end{figure*}

This Appendix details the smoothing algorithm used for spectral fitting of all quasars in this work. A visual demonstration of the procedure is shown in Figure~\ref{fig:smoothing}.

The challenges this algorithm has to overcome are twofold. First, we need to compute a fit of the observed flux which discards all absorption features, since these are not part of the \textit{intrinsic} spectra we aim to extract correlations from. Second, we need to capture the full strength of emission line peaks in the spectra without damping them, since these are used to establish the spectral correlations. Based on these two challenges, we developed the following smoothing algorithm which successfully discards all reasonably narrow absorption features while maintaining the full strength of even the strongest and sharpest emission lines, especially the Ly$\alpha$ peak.

We first compute a running median with a bin size of 50 data points in order to capture the main continuum and the overall spectral shape of the spectrum. This curve then acts like a median border (Figure~\ref{fig:smoothing} (a), yellow), above which we then perform a peak-finding algorithm using the SciPy Python library \citep{scipy} to find local maxima in the spectrum (Figure~\ref{fig:smoothing} (a), red points).

We then interpolate the peaks to construct an upper envelope of the spectrum (Figure~\ref{fig:smoothing} (b), red). This envelope is then subtracted from the raw data points, thus linearizing our data into residuals (Figure~\ref{fig:smoothing} (c), black). We then apply the RANSAC regressor algorithm \citep{fischler1981random} from the Scikit-Learn Python package \citep{2012arXiv1201.0490P} on the residuals, which fits a linear function to the data and masks all data points as either inliers or outliers. By further considering only inlying data points (Figure~\ref{fig:smoothing} (c), green), we thus reject most absorption features in the spectrum.

Finally, the data points which are flagged as inliers by RANSAC are interpolated and smoothed by computing a running median with a bin size of 20, thus creating the final flux fit of each spectrum (Figure~\ref{fig:smoothing} (d), cyan). This algorithm will be made public in near future.

\section{Analysis of redshift calibration systematics}\label{z_app}

This section discusses and analyses an important potential source of systematic errors, namely the uncertainty in the QSO redshifts, both the low-redshift ones (i.e. training set SDSS quasars) and the high-redshift ones (i.e. ULAS~J1120+0641 and ULAS~J1342+0928). Moreover, while the SDSS redshifts have been calculated based on the broad UV emission lines, the redshifts of the two $z>7$ QSOs come from the sub-milimeter emission from their host galaxies \citep{2017ApJ...837..146V,2017ApJ...851L...8V,2018Natur.553..473B}. We analyse how resampling the redshifts of the SDSS QSOs in the training set influences the performance and predictions of \textsc{QSANNdRA}, and we also investigate how changing the redshifts of the high-redshift QSOs influences the predicted continua and hence neutral fractions. As a final test, we recalibrate both the low-redshift and the high-redshift spectra based on the redshift coming from the Mg\Romannum{2} line, since it has been shown to be the least affected by systematic shifts \citep{2010MNRAS.405.2302H,2016ApJ...831....7S}.

\begin{figure}
    \centering
    \includegraphics[width=\columnwidth]{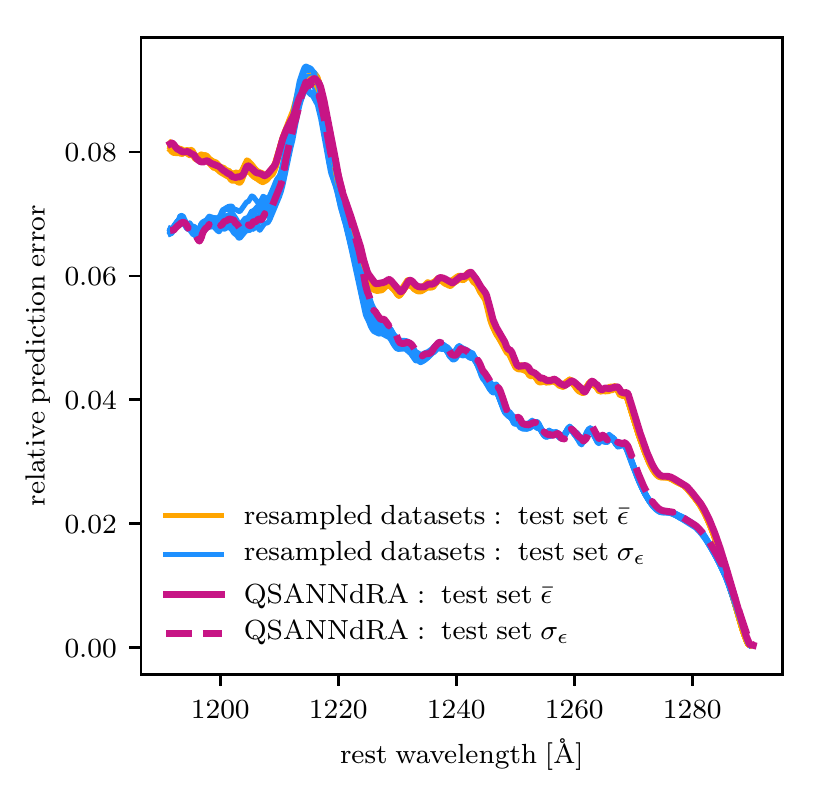}
    \caption{Wavelenght-dependence of the test set performance of our model based on the 10 resampled datasets displayed as $\bar{\epsilon}$ (orange) and $\sigma_\mathrm{\epsilon}$ (blue). Both of these metrics are almost identical to the baseline performance from the main text (magenta), hence showing that redshift calibration of the training set quasars does not influence the performance of our model.}.
    \label{fig:resample_err}
\end{figure}

\begin{figure}
    \centering
    \includegraphics[]{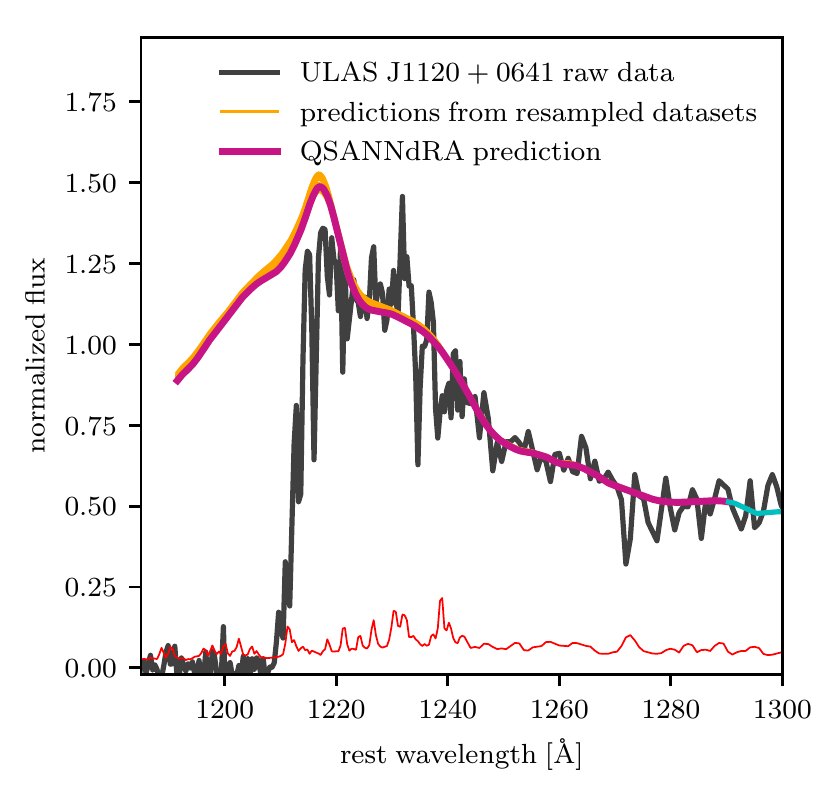}
    \includegraphics[]{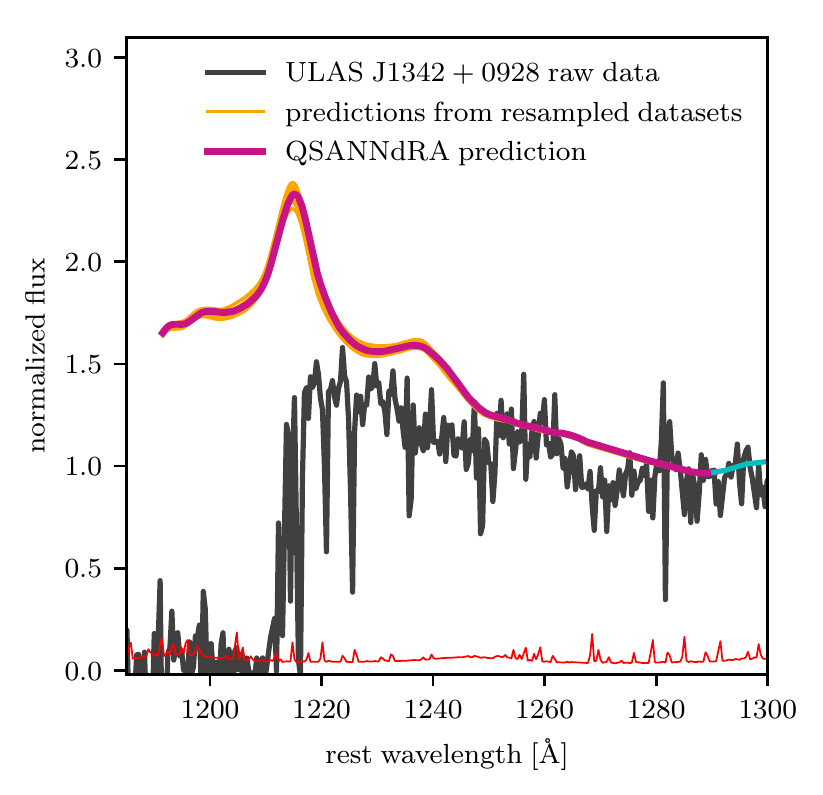}
    \caption{A comparison of the 10 new predictions (orange curves) based on the resampled data sets and \textsc{QSANNdRA}'s baseline prediction for ULAS~J1120+0641 (top) and ULAS~J1342+0928 (bottom). All predictions almost completely overlap, the most significant exception being the Ly$\alpha$ peak where the variance is the greatest, yet still considerably small. Even though this influences the resultant neutral fraction constraint, our model seems to be robust against changes in the redshift calibration of the SDSS training set QSOs.}
    \label{fig:J1120_res}
\end{figure}

The first part of our analysis involves investigating how changing the redshifts of the SDSS QSOs and retraining our model changes the performance and predictions for the two high-redshift QSOs. We do this by resampling our training set QSOs 10 times to create 10 different training sets. For each new data set, noting that each SDSS QSO has a redshift uncertainty \texttt{Z{\_}ERR} assigned to it which defines the variance of its redshift distribution, we randomly sample the redshift of each QSO along this distribution and calibrate the observed wavelengths to the correct rest-frame accordingly. In doing this, we model the redshift distribution of each QSO as Gaussian with a mean of \texttt{Z} and a sigma of \texttt{Z{\_}ERR}. With these 10 resampled training sets, we proceed in the exact same way as reported in the main text, i.e. we perform a 10-fold cross-validation to clean up the training sets, we train \textsc{QSANNdRA} on these new data sets, and we predict high-redshift quasar continua along with neutral fraction constraints.

After training, we evaluate the performance of our model on the test set of low-redshift QSOs by means of the mean absolute prediction error $\bar{\epsilon}$ and its standard deviation $\sigma_\mathrm{\epsilon}$. Figure~\ref{fig:resample_err} displays \textsc{QSANNdRA}'s performance on each of the 10 resampled data sets as $\bar{\epsilon}$ in orange and $\sigma_\mathrm{\epsilon}$ in blue, along with its performance from the main text in magenta. As can be clearly seen, the differences in the performances are marginal and hence we conclude that redshift calibration does not significantly influence the performance our model is able to achieve, if the errors are distributed randomly.  Note that this may change if the redshifts are systematically biased in any way for certain types of QSOs.  Given the complexity of our NNs, we do expect that our training should be able to account for some of this systematic bias if it is to due to the redshift estimation from specific emission lines.

With the 10 new trained models, we reconstruct the blue-side continua for both ULAS~J1120+0641 and ULAS~J1342+0928. Figure~\ref{fig:J1120_res} shows all 10 resultant predictions for ULAS~J1120+0641 (top) and ULAS~J1342+0928 (bottom), respectively, as a set of orange lines and the predictions given in the main text in magenta for comparison. In each case, all resultant predictions from the resampled data sets almost completely overlap with our baseline prediction, which hints at a very marginal influence of redshift calibration on the predicted continua themselves. However, it should be noted that the largest spread in predictions occurs at the Ly$\alpha$ peak in both cases, which can in turn influence the predicted neutral fraction constraints.

\begin{figure}
    \centering
    \includegraphics[width=\columnwidth]{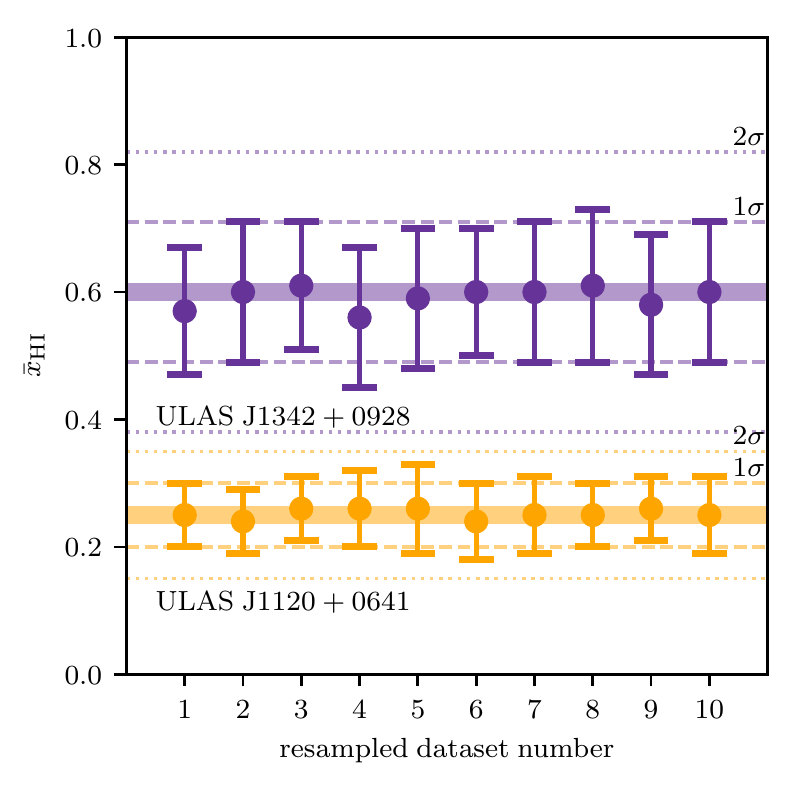}
    \caption{A comparison of the predicted neutral fractions based on the 10 resampled data sets for ULAS~J1120+0641 (orange) and ULAS~J1342+0928 (purple), to the baseline predictions from the main text (thick horizontal lines). All neutral fraction constraints based on the resampled data sets are consistent with each other as well as with the baseline constraints $\bar{x}_\mathrm{H\Romannum{1}} = 0.25^{+0.05}_{-0.05}$ for ULAS J1120+0641 and $\bar{x}_\mathrm{H\Romannum{1}} = 0.60^{+0.11}_{-0.11}$ for ULAS J1342+0928.}
    \label{fig:resample_xHI}
\end{figure}

The resultant neutral fractions computed by the 10 retrained models are displayed in Figure~\ref{fig:resample_xHI} along with their corresponding 68\% bounds for both ULAS~J1120+0641 (orange) and ULAS~J1342+0928 (purple). We also show the original predictions from the main text as gray horizontal lines as well as their corresponding 1$\sigma$ (dashed) and 2$\sigma$ (dotted) bounds. All 10 predictions are consistent with each other and also with the fiducial prediction in each case. 

\begin{figure}
    \centering
    \includegraphics[]{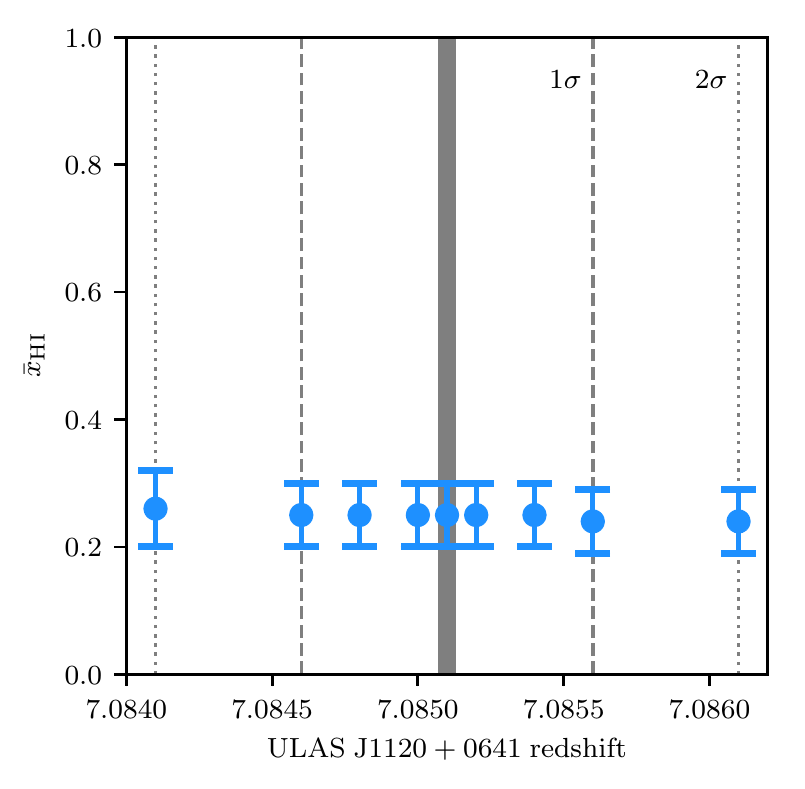}
    \includegraphics[]{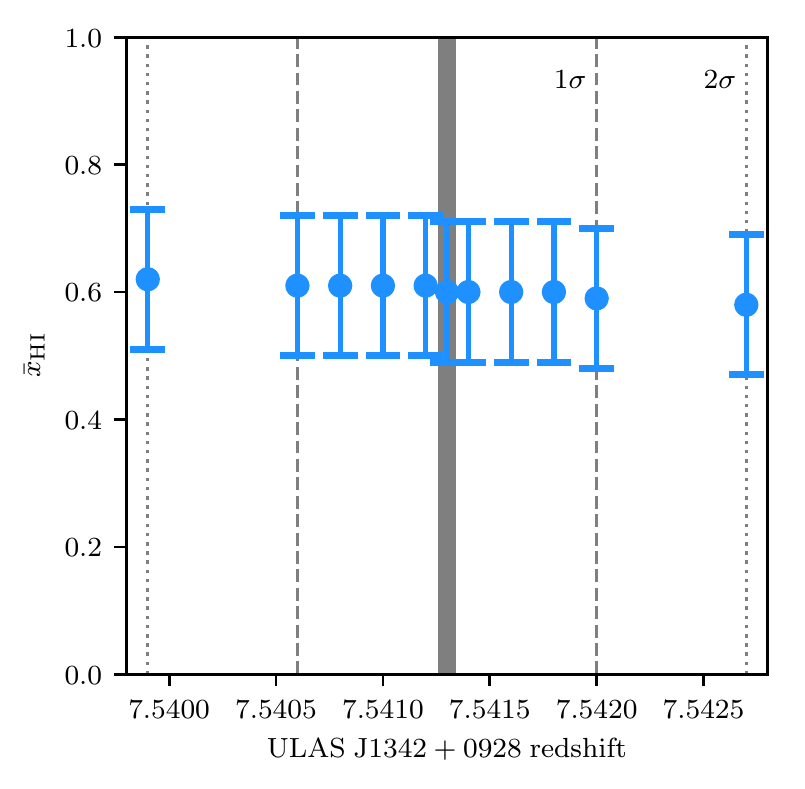}
    \caption{Dependence of the estimated neutral fraction $\bar{x}_\mathrm{H\Romannum{1}}$ predicted by \textsc{QSANNdRA} for ULAS~J1120+0641 (top) and ULAS~J1342+0928 (bottom) on the redshift of the QSO.  The vertical, solid gray line depict the reported redshifts, while the dashed and dotted lines represent the 1$\sigma$ and 2$\sigma$ uncertainties, respectively.  The error bars on the blue points represent the 68\% confidence interval on the estimate of the neutral fraction.}
    \label{fig:J1120_reds}
\end{figure}

\begin{figure*}
   \centering
   \includegraphics[width=\linewidth,trim={0cm 0.7cm 0cm 0cm},clip]{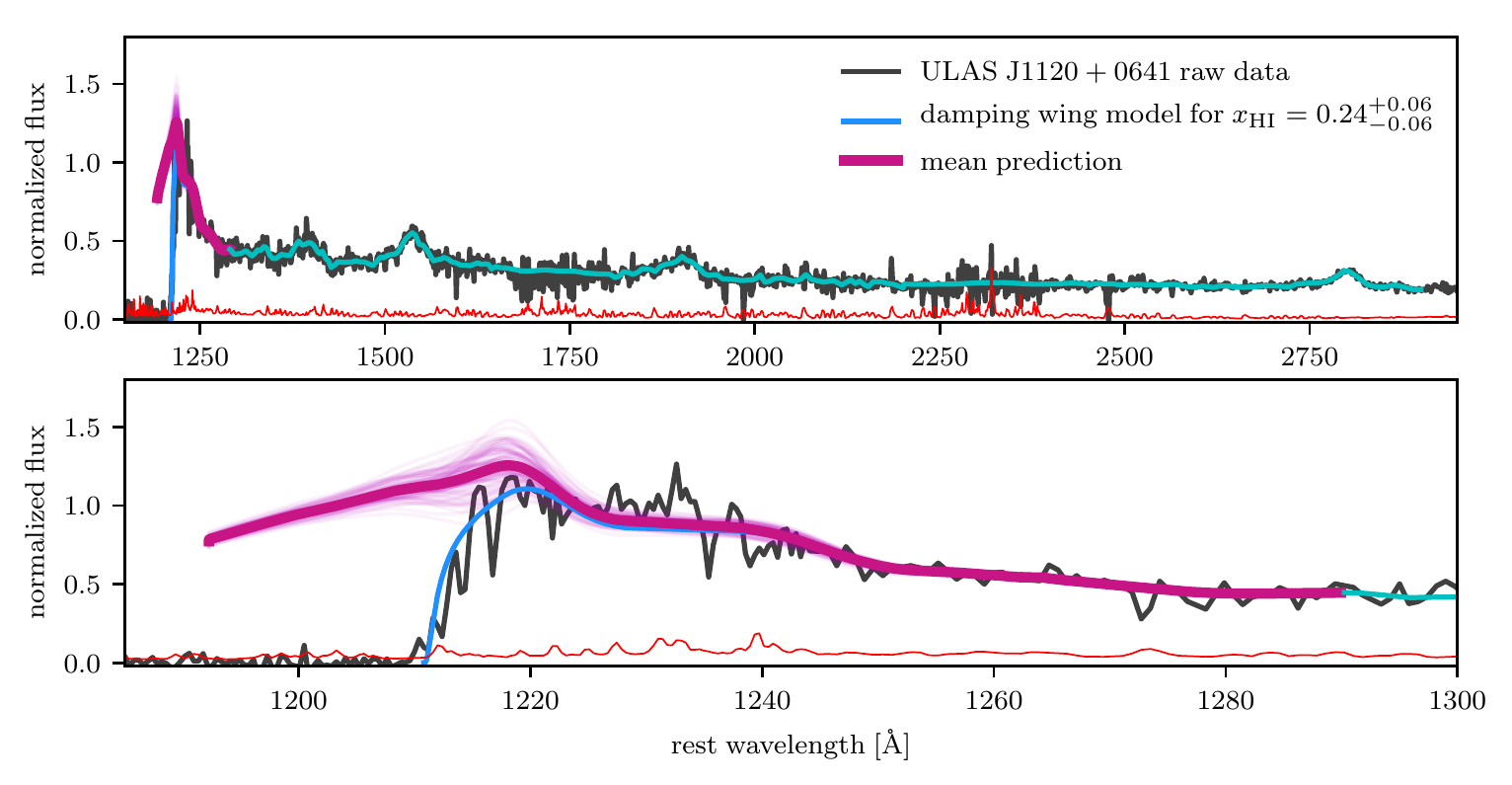}
   \includegraphics[width=\linewidth]{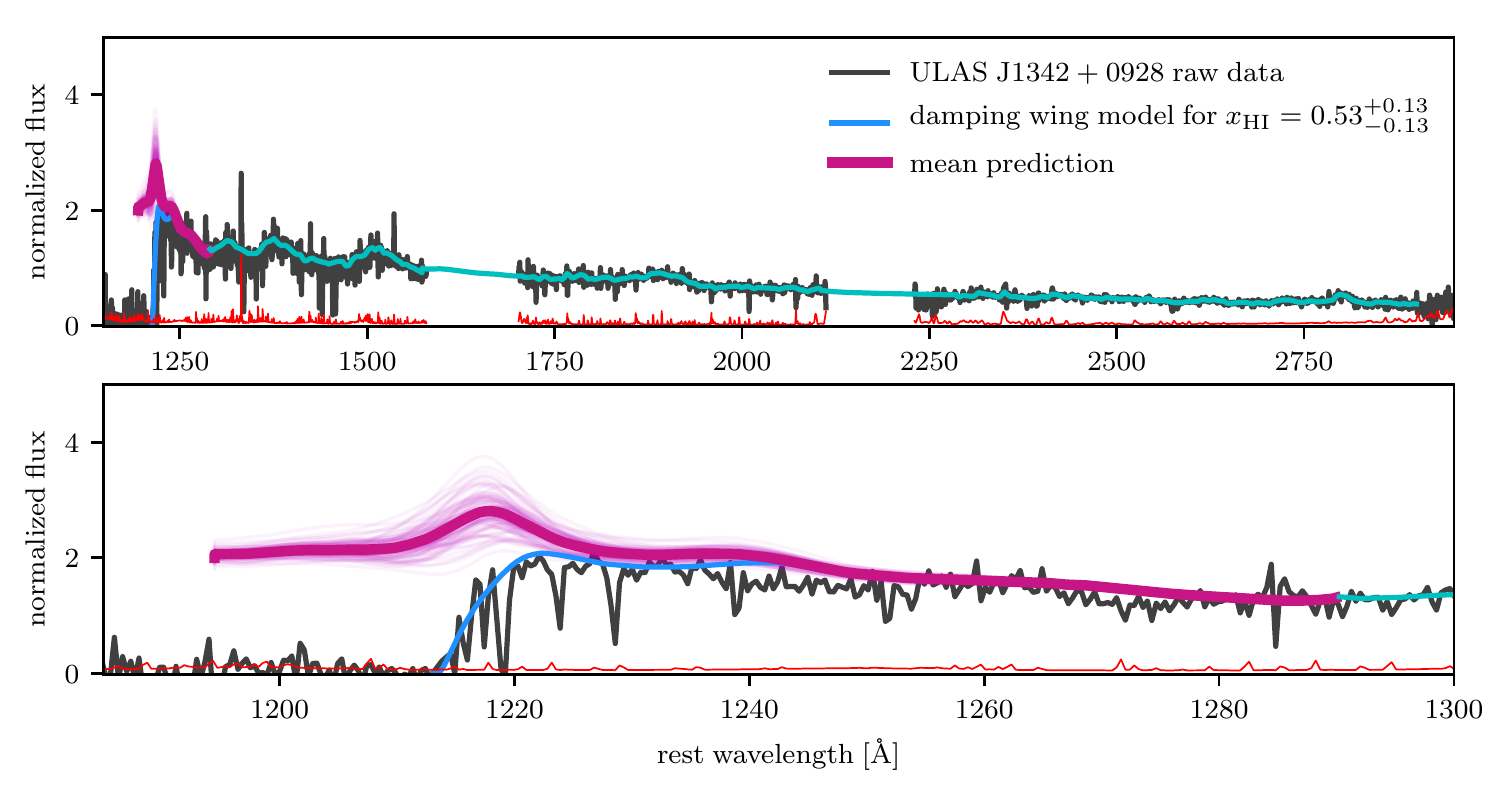}
   \caption{The reconstructed spectrum of ULAS~J1120+0641 (top two panels) and ULAS~J1342+0928 (bottom two panels) based on a test model recalibrated according to the Mg\Romannum{2} emission line redshift. The bottom panel for each quasar shows a close-up view of the Ly$\alpha$ region. The raw data points and their uncertainties are shown in gray and red, respectively. The cyan curve represents our fit of the red-side spectrum. The thin light magenta lines show the individual predictions from the 100 retrained NNs in the committee, while the thick magenta line shows the weighted average of these predictions at each wavelength. The damping wing model is shown in blue and corresponds to $\bar{x}_\mathrm{H\Romannum{1}} = 0.24$ for ULAS~J1220+0641 and $\bar{x}_\mathrm{H\Romannum{1}} = 0.53$ for ULAS~J1342+0928, which was calculated as the weighted average of optimal neutral fractions corresponding to the individual predictions of the 100 networks within the committee. Note that while the predictions are based on estimates of $z_{\mathrm{z_Mg\Romannum{2}}}$ from the fitted spectrum of the QSOs, the damping wing analysis was performed for the much more accurate redshifts, i.e. $z=7.0851$ and $z=7.5413$, respectively.}
   \label{fig:z_Mg}
\end{figure*}

The second part of this analysis investigates how \textsc{QSANNdRA}'s predictions change as we vary the redshift of the two high-redshift QSOs. \cite{2017ApJ...837..146V} reported a redshift of $7.0851^{+0.0005}_{-0.0005}$ for ULAS~J1120+0641, and \cite{2017ApJ...851L...8V} and \cite{2018Natur.553..473B} reported a redshift of $7.5413^{+0.0007}_{-0.0007}$ for ULAS~J1342+0928. Hence, we rerun our fiducial model on these two QSOs again, each time changing the redshift based on which calibration from observed to rest-frame wavelengths was carried out. Figure~\ref{fig:J1120_reds} shows the resultant neutral fractions as blue data points with errorbars corresponding to 68\% bounds predicted by \textsc{QSANNdRA}. The reported redshift and 1$\sigma$ and 2$\sigma$ bounds are shown as dashed and dotted gray lines, respectively.

As can be observed, the relationship between the redshift of both high-redshift QSOs and the estimated $\bar{x}_\mathrm{H\Romannum{1}}$ seems to be approximately linear and is likely to be the consequence of all the spectral features being translated along the wavelength space, thus changing the values of red-side PCA coefficients that \textsc{QSANNdRA} is basing its predictions on. Within the 2$\sigma$ uncertainty in redshift quoted for the both QSOs, there is virtually no change in the estimated neutral fraction.

As the final test, we used redshifts based on the Mg\Romannum{2} line to recalibrate both the SDSS and $z>7$ spectra in a unified fashion and then retrained our model to predict the high-redshift continua and neutral fraction constraints. This particular line was chosen due to its minimal systematic shifts \citep{2010MNRAS.405.2302H,2016ApJ...831....7S}. For the low-redshift QSOs, we used the Mg\Romannum{2} redshift from the SDSS pipelines, while for the high-redshift quasars, we estimated the corresponding redshift based on the Mg\Romannum{2} peak wavelength of the fitted continuum. However, since the sub-milimeter redshifts ($z=7.0851$ and $z=7.5413$) for the high-redshift QSOs are physically much more accurate than an estimate from the Mg\Romannum{2} emission line, we perform the damping wing analysis in the rest frame of these quasars defined by $z$.

In Figure~\ref{fig:z_Mg} we show the resultant predictions and neutral fraction constraints for both $z>7$ QSOs. We observe that there is a minimal change in the shape of both predicted continua as well as the strength of the Ly$\alpha$ peak. Note that the y-axis in both plots has been normalized with respect to the fitted flux at 1290{\AA}, which corresponds to a different value than that in the main text. In addition, the new neutral fraction constraints, namely $\bar{x}_\mathrm{H\Romannum{1}} = 0.24^{+0.06}_{-0.06}$ at $z=7.0851$ and $\bar{x}_\mathrm{H\Romannum{1}} = 0.53^{+0.13}_{-0.13}$ at $z=7.5413$, are consistent with the constraints from the main text ($\bar{x}_\mathrm{H\Romannum{1}} = 0.25^{+0.05}_{-0.05}$ and $\bar{x}_\mathrm{H\Romannum{1}} = 0.60^{+0.11}_{-0.11}$, respectively). We therefore conclude that the systematics due to different redshift calibrations do not cause significant errors in our model.

\vspace{-15pt}
\section{Dependence of neutral fraction constraints on \texorpdfstring{$z_\mathrm{N}$}{zN}}\label{z_n_app}

Here we report an analysis of how the exact choice of the $z_\mathrm{N}$ parameter in the damping-wing model \citep{1998ApJ...501...15M}, which defines the redshift by which the IGM is completely reionized, impacts the predicted neutral fraction constraints for the two high-redshift quasars, ULAS~J1120+0641 and ULAS~J1342+0928.

\begin{figure}
    \centering
    \includegraphics[width=\columnwidth]{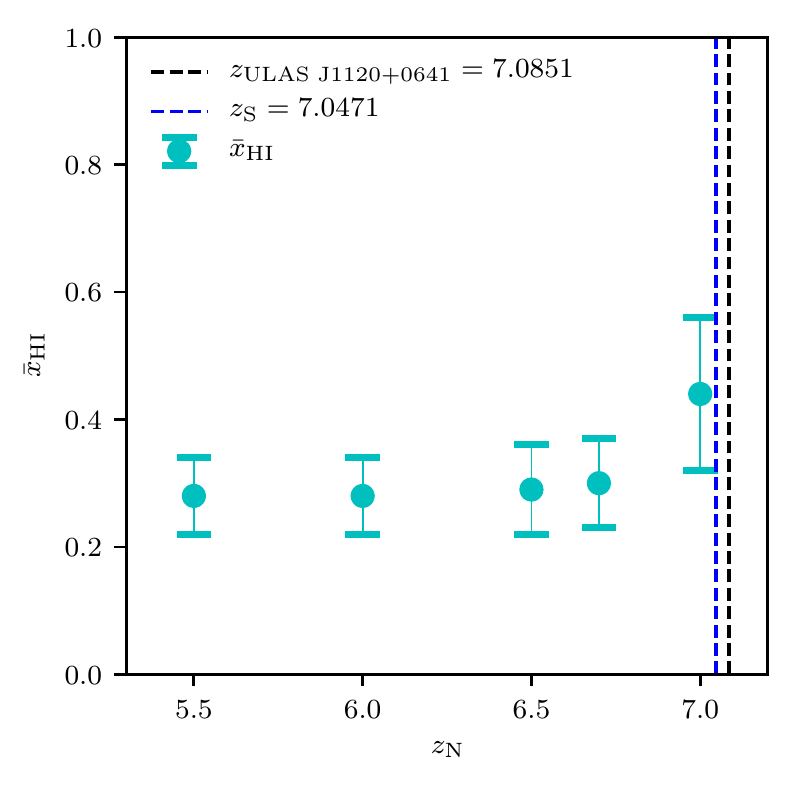}
    \includegraphics[width=\columnwidth]{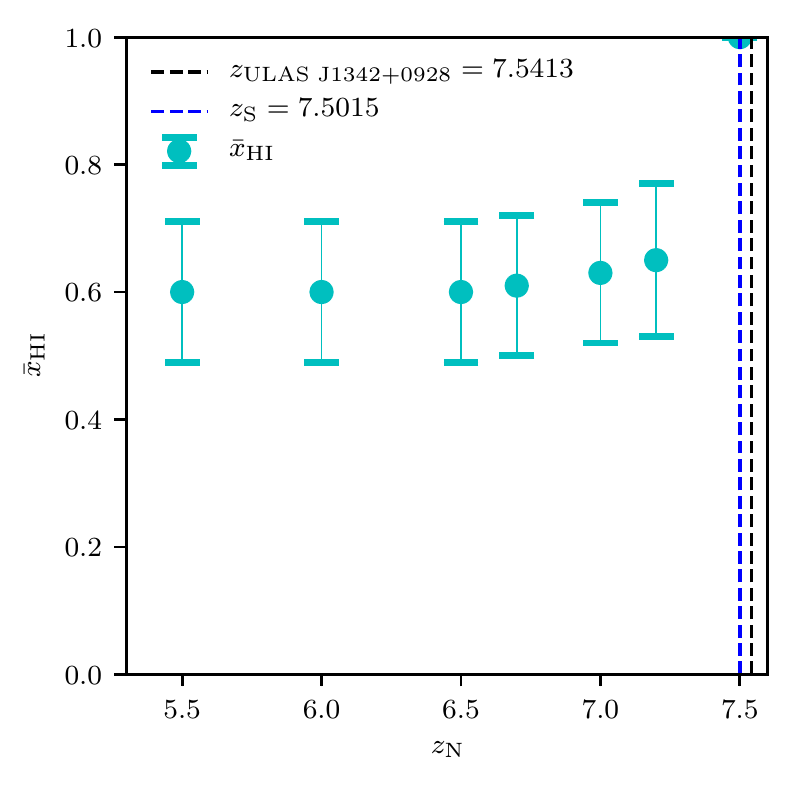}
    \caption{Dependence of the predicted neutral fraction $\bar{x}_\mathrm{H\Romannum{1}}$ for ULAS~J1120+0641 (top) and for ULAS~J1342+0928 (bottom) on the value of the redshift $z_\mathrm{N}$, by which we assume the IGM to be fully ionized. We confirm that this dependence is very weak unless $z_\mathrm{N}$ nears the redshift corresponding to the end of the quasar's proximity zone in the model used in the main text.}
    \label{fig:J1120_z_N}
\end{figure}

In this analysis, we use the fiducial \textsc{QSANNdRA} algorithm as described in the main text. We re-run the prediction algorithm multiple times, each time with a different value of $z_\mathrm{N}$ ranging from $z=5.5$ up to the redshift of the particular quasar while keeping everything else constant. Figure~\ref{fig:J1120_z_N} shows the resultant neutral fractions and their 68\% bounds for ULAS~J1120+0641 and ULAS~J1342+0928.

In each case, we confirm that the exact value of $z_\mathrm{N}$ does not have a significant impact on the predictions provided it is not nearing the redshift corresponding to the end of the quasar's near zone $z_\mathrm{S}$. This makes physical sense since the difference between $z_\mathrm{N}$ and $z_\mathrm{S}$ constrains the distance range over which the observed damping by neutral hydrogen in the IGM needs to happen. If this range gets extremely small, the neutral fraction needed to reconstruct the observed damping-wing must increase, which is what we see in the last few data points nearing the QSO redshifts in Figure~\ref{fig:J1120_z_N}.

Overall, these results confirm that our choice of $z_\mathrm{N} = 6$ is not particularly important and that our main results can be generalised well to values of $z_\mathrm{N}\lesssim 6.5$.


\bsp	
\label{lastpage}
\end{document}